\documentclass{aa}  

\usepackage{graphicx}
\usepackage{txfonts}
\usepackage{longtable}
\begin{document}

   \title{The SOPHIE search for northern extrasolar planets\thanks{Based on observations collected with the SOPHIE spectrograph on the 1.93-m telescope at Observatoire de Haute-Provence (CNRS), France by the SOPHIE Consortium.}}

   \subtitle{XIII. Two planets around M-dwarfs Gl617A and Gl96}

   \author{M. J. Hobson
          \inst{\ref{i:lam}}
          \and
          R. F. D\'iaz\inst{\ref{i:uba},\ref{i:conicet}}
          \and
          X. Delfosse\inst{\ref{i:grenoble}}
          \and
          N. Astudillo-Defru\inst{\ref{i:geneve}}
          \and
          I. Boisse\inst{\ref{i:lam}}
          \and
          F. Bouchy\inst{\ref{i:geneve}}
          \and
          X. Bonfils\inst{\ref{i:grenoble}}
          \and
          T. Forveille\inst{\ref{i:grenoble}}
          \and
          N. Hara\inst{\ref{i:geneve}, \ref{i:ASD}}
          \and
          L. Arnold\inst{\ref{i:OHP}}
          \and
          S. Borgniet\inst{\ref{i:grenoble}}
          \and
          V. Bourrier\inst{\ref{i:geneve}}
          \and
          B. Brugger\inst{\ref{i:lam}}
          \and
          N. Cabrera\inst{\ref{i:grenoble}}
          \and
          B. Courcol\inst{\ref{i:lam}}
          \and
          S. Dalal\inst{\ref{i:paris}}
          \and
          M. Deleuil\inst{\ref{i:lam}}
          \and
          O. Demangeon\inst{\ref{i:porto}}
          \and
          X. Dumusque\inst{\ref{i:geneve}}
          \and
          D.Ehrenreich\inst{\ref{i:geneve}}
          \and
          G. Hébrard\inst{\ref{i:paris}, \ref{i:OHP}}
          \and
          F. Kiefer\inst{\ref{i:paris}}
          \and
          T. Lopez\inst{\ref{i:lam}}
          \and
          L. Mignon\inst{\ref{i:grenoble}}
          \and
          G. Montagnier\inst{\ref{i:paris}, \ref{i:OHP}}
          \and
          O. Mousis\inst{\ref{i:lam}}
          \and
          C. Moutou\inst{\ref{i:lam},\ref{i:cfht}}
          \and
          F. Pepe\inst{\ref{i:geneve}}
          \and
          J. Rey\inst{\ref{i:geneve}}
          \and
          A. Santerne\inst{\ref{i:lam}}
          \and
          N. Santos\inst{\ref{i:porto},\ref{i:porto2}}
          \and
          M. Stalport\inst{\ref{i:geneve}}
          \and
          D. Ségransan\inst{\ref{i:geneve}}
          \and
          S. Udry\inst{\ref{i:geneve}}
          \and
          P.A. Wilson\inst{\ref{i:leiden},\ref{i:paris}}
          }

   \institute{Aix Marseille Université, CNRS, Laboratoire d’Astrophysique de Marseille UMR 7326, 13388 Marseille cedex 13, France\\
              \email{melissa.hobson@lam.fr}\label{i:lam}
            \and
         Universidad de Buenos Aires, Facultad de Ciencias Exactas y Naturales. Buenos Aires, Argentina\label{i:uba}
         \and
         CONICET - Universidad de Buenos Aires. Instituto de Astronomía y Física del Espacio (IAFE). Buenos Aires, Argentina\label{i:conicet}
         \and
          Univ. Grenoble Alpes, CNRS, IPAG, 38000 Grenoble, France\label{i:grenoble} 
         \and
         Observatoire Astronomique de l’Université de Genève, 51 Chemin des Maillettes, 1290 Versoix, Switzerland\label{i:geneve}
        \and
        ASD/IMCCE, CNRS-UMR8028, Observatoire de Paris,  PSL, UPMC, 77 Avenue Denfert-Rochereau, 75014 Paris, France \label{i:ASD}
        \and
         Observatoire de Haute Provence, CNRS, Aix Marseille Université, Institut Pythéas UMS 3470, 04870 Saint-Michel-l’Observatoire, France\label{i:OHP}
         \and
         Institut d’Astrophysique de Paris, UMR7095 CNRS, Université Pierre \& Marie Curie, 98bis boulevard Arago, 75014 Paris, France\label{i:paris}
         \and
         Instituto de Astrofísica e Ciências do Espaço, Universidade do Porto, CAUP, Rua das Estrelas, 4150-762 Porto, Portugal\label{i:porto}
         \and
         Departamento de Física e Astronomia, Faculdade de Ciências, Universidade do Porto, Rua do Campo Alegre, 4169-007 Porto, Portugal\label{i:porto2}
         \and
         Canada-France-Hawaii Telescope Corporation, 65-1238 Mamalahoa Hwy, Kamuela, HI 96743, USA\label{i:cfht}
         \and
         Leiden Observatory, Leiden University, Postbus 9513, 2300 RA Leiden, The
        Netherlands\label{i:leiden}
        }

   \date{Received January 30, 2018, accepted June 27, 2018}

% \abstract{}{}{}{}{} 
% 5 {} token are mandatory
 
  \abstract
   {We report the detection of two exoplanets and a further tentative candidate around the M-dwarf stars Gl96 and Gl617A, based on radial velocity measurements obtained with the SOPHIE spectrograph at the Observatoire de Haute Provence. Both stars were observed in the context of the SOPHIE exoplanet consortium's dedicated M-dwarf subprogramme, which aims to detect exoplanets around nearby M-dwarf stars through a systematic survey. For Gl96, we present the discovery of a new exoplanet at 73.9 d with a minimum mass of 19.66 earth masses. Gl96 b has an eccentricity of 0.44, placing it among the most eccentric planets orbiting M stars. For Gl617A we independently confirm a recently reported exoplanet at 86.7 d with a minimum mass of 31.29 earth masses. Both Gl96 b and Gl617A b are potentially within the habitable zone, though Gl96 b's high eccentricity may take it too close to the star at periapsis.}


   \maketitle
%
%-------------------------------------------------------------------

\section{Introduction}
\label{s:int}

M-dwarf stars are both interesting and promising targets for exoplanet hunts. They are the most common stars in the Galaxy, and studies suggest their planet occurrence rates are high (e.g. \citealt{Bonfils}, \citealt{Dressing}). Moreover, they are interesting candidates for habitable planet searches. Their relatively small masses ($0.07 - 0.6 M_\odot$, \citealt{Reid}) mean that small planets will still induce detectable signals, while their faintness compared to G-type stars means the habitable zone is located closer to the stars. Hence, low-mass short-period habitable planets are easier to detect for M-dwarfs than for sun-like stars - e.g. Gl 667C c, the first habitable-zone Earth-size planet around an M-dwarf \citep{Delfosse13}; LHS 1140 b, one of the most recently detected ones, orbiting the brightest M-dwarf with a transiting planet in the habitable zone \citep{Dittmann}; TRAPPIST-1 e, f, g, three potentially habitable telluric planets in a seven-planet system \citep{Gillon}; GJ 273, with two super-Earths one of which is in the habitable zone \citep{Astudillo17b}; K2-18b, a transiting habitable-zone planet whose mass was characterised by RV follow-up, revealing a density that may correspond to a rocky planet with extended atmosphere or to a water world \citep{Foreman, Cloutier}.

Currently, 146 exoplanets around main sequence M-dwarf stars are known, compared to 997 planets around FGK stars\footnote{Retrieved on Oct 11, 2017, from \citealt{exoplanet}, considering only the stars for which spectral type is reported in the catalogue, and filtering out those not on the main sequence.}. However, this number is expected to grow, as several current or near-future projects have M-dwarf stars as part of their (or their sole) primary targets, e.g. SPIRou \citep{Artigau}, TESS (NASA mission, launch 2018, \citealt{Ricker}), TRAPPIST (e.g. \citealt{Gillon}), CARMENES (e.g. \citealt{Quirrenbach}), HADES (e.g. \citealt{Affer}), NIRPS \citep{Bouchy17}, ExTrA \citep{Bonfils15}. Given the relatively low number of detected exoplanets around M-dwarfs, each new detection provides valuable information on the population that can be used to refine observing strategies.

Since 2006, the SOPHIE exoplanet consortium has been carrying out several planet-hunting programmes using the SOPHIE spectrograph at the Observatoire de Haute-Provence \citep{Bouchy09}. Sub-programme 3, or SP3, consists of a systematic survey of nearby M-dwarfs which aims at detecting habitable SuperEarths and Neptunes, constraining the statistics of planets around M-dwarfs, and finding potentially transiting companions. A complete description of the SP3 is beyond the scope of this paper; the programme will be presented in detail in a forthcoming paper.

In this work, we report the results of the SP3 study of two M-dwarfs, Gl96 and Gl617A. Section \ref{s:obs} presents the observations. In Section \ref{s:data} we describe the analysis of the data. Our results are presented in Section \ref{s:res}, together with an analysis of the Hipparcos photometry of these stars in Section \ref{s:phot}, and discussed in Section \ref{s:disc}.

%-------------------------------------------------------------------

\section{Observations}
\label{s:obs}

Gl96 and Gl617A were observed with the SOPHIE spectrograph as part of the SOPHIE consortium search for exoplanets around M-dwarfs. SOPHIE is a fibre-fed, environmentally stabilised, cross-dispersed echelle spectrograph mounted on the 193 cm telescope at the Observatoire de Haute-Provence \citep{Perruchot08}. In 2011, SOPHIE was upgraded by inserting an octagonal-section fibre in the fibre link \citep{Perruchot11, Bouchy13}. The upgraded spectrograph is known as SOPHIE+.

The observations presented here were performed after the 2011 upgrade, using the high-resolution (HR) mode of the spectrograph, for a resolving power of $\lambda/\delta\lambda\approx75000$. The SOPHIE spectrograph provides two modes for fibre B: \textit{thosimult} mode, in which a simultaneous ThAr calibration is performed and is used to trace the spectrograph drift during the night, and \textit{objAB} mode, in which fibre B is used to monitor the sky brightness. The choice of modes for the SP3 targets depends on the brightness of the stars (with a limit at V=9). Gl96, which is above this limit, was observed in the \textit{objAB} mode to control for possible moonlight contamination and avoid any potential ThAr contamination, with a calibration lamp spectrum obtained immediately before each observation to monitor potential drifts. Gl617A, which is brighter, was observed in the \textit{thosimult} mode, since it is bright enough that the contamination by moonlight can be neglected, and contamination from the ThAr spectrum on fibre B is negligible.

Observations were gathered between 2011 and 2017. In total, 79 spectra were obtained for Gl96 with SOPHIE+, with a median exposure time of 1800s and a median SNR at 550 nm of 82.7. 163 spectra were obtained for Gl617A, with a median exposure time of 900s and a median SNR at 550 nm of 84.1.

The SOPHIE pipeline \citep{Bouchy09} was used to reduce and extract the spectra. The spectra were then cross-correlated with an M3 stellar spectral mask in order to obtain the cross-correlation functions (CCFs), from which radial velocities (RV), FWHM, contrast, and bisectors were measured. The mask was built from the median of a large number of spectra of Gl581 obtained with HARPS (La Silla, ESO) and degraded to the SOPHIE spectral resolution.

For Gl617A, we removed 9 spectra with SNR<35. For Gl96, we removed a total of 8 observations: 2 spectra following the same SNR criteria as for Gl617A, and 6 spectra due to moon contamination. 

%--------------------------------------------------------------------

\section{Data Analysis}
\label{s:data}

Radial velocities for stabilised spectrographs with ThAr-derived wavelength calibration have usually been obtained by the CCF method, in which the spectra are correlated with a weighted binary mask (see \citealt{Queloz95} and \citealt{Pepe02} for full descriptions of the method). Although the CCF method is very effective for FGK stars, which have strong spectral lines and a well defined continuum, this is not the case for M-dwarfs, where numerous overlapping molecular bands complicate the continuum determination. For these stars, the use of these binary masks which target only clearly defined lines underutilises the Doppler information present in the spectrum. Therefore, other methods have been developed to better exploit this information, such as template-matching using a true stellar template (as done, e.g., by the HARPS-TERRA code of \citealt{Anglada12}). Template-matching also allows a more precise removal of the telluric lines, and of any parts of the spectra that are not compliant with the template or that have no spectral information. In this work, we make use of an algorithm developed by N. Astudillo-Defru (\citealt{Astudillo15}, \citealt{Astudillo17b}) for this purpose; it constructs stellar and telluric templates from the observed spectra, discards the telluric-contaminated zones, and derives the radial velocity by $\chi^2$ minimisation, using the RV determined by the CCF method as a first guess. We applied this method to all the SOPHIE+ spectra.

Before employing the template-matching procedure, we performed a correction for the charge transfer inefficiency (CTI) effect, following the characterisation of \cite{Bouchy09b}. Once the RVs were determined, we added a further correction for the instrumental drift using the simultaneous ThAr calibration (in the case of Gl617A) or an interpolation between ThAr calibrations performed before and after the exposure (for Gl96). 

\begin{figure}
\centering
\includegraphics[width=\hsize]{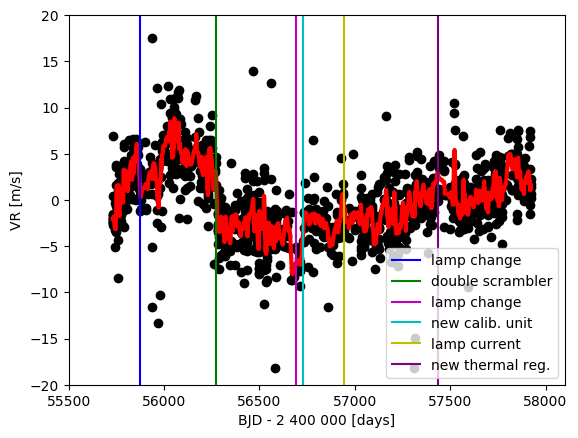}
\caption{Zero-point drift correction (red line) and radial velocities of the 14 stars used to construct it (black dots). The vertical lines indicate identified jumps and their causes (see Section \ref{s:data} for details). The zero-point correction spans 6 years; it has a dispersion of 3.15 m/s, and a peak-to-peak variation of 16.9 m/s.}
\label{master}
\end{figure}

The SOPHIE+ spectrograph also presents long-term variations of the zero-point, an effect described in \cite{Courcol15}. In order to correct it, the authors iteratively construct a master RV time series from RV constant stars. This master series is used to correct the measurements of a given target by subtracting the velocity in the master series, interpolated at the dates of observation. We followed the procedure described in \cite{Courcol15} to construct an analogous master for the SP3 programme, using the ensemble of SP3 targets with at least 10 SOPHIE+ observations plus the four 'super-constants' from the SP1 programme (defined by \citealt{Courcol15}) as a starting point. We chose to use primarily SP3 targets when constructing our master correction in order to mitigate any potential bias or offset due to differing spectral types. The final master employs 14 stars: the SP1 (G-type) constants HD185144, HD9407, HD22154, and HD89269A; three M-dwarfs which are systematically observed for all observation seasons and considered as our SP3 constants, Gl411, Gl514, and Gl686; and the additional SP3 stars G239-25, Gl133, Gl15A, Gl436, Gl521, Gl694, and Gl728 (all additional stars used for the constant correction have a corrected rms after the first iteration lower than 3 m/s, as defined by \citealt{Courcol15}). We also tested a master constructed using only the SP3 targets, but found the resulting RVs had higher dispersion (around 1.5 m/s higher on average). Figure \ref{master} shows the master used for the zero-point drift correction and the RVs from which it was derived.

The zero-point drift correction clearly reflects several instrument modifications: a jump at 55872 (6/11/2011) corresponding to a ThAr calibration lamp change; a jump at 56274 (12/12/2012) following the installation of octagonal fibres after the double scrambler; a second ThAr lamp-change related jump at 56690 (01/02/2014); a jump at 56730 (13/03/2014) after the installation of a new calibration unit; a jump at 56941 (10/10/2014) due to a change in the current of the ThAr calibration lamp; and a jump at 57435 corresponding to the installation of a new thermal regulation. These events are indicated in Figure \ref{master}. An additional long-term drift is also seen.

The first two events described were also noted by \cite{Courcol15}. We do not, however, find the jump at 56775 which the authors correlated at the time to the recoating of the secondary mirror (including when we regarded a test master constructed using only the template-matching derived RVs of the SP1 super-constants employed by \citealt{Courcol15}). We have identified the actual cause of the jump at 56775 as an instability of the wavelength solution on blue spectral orders, whose weight changed after mirror recoating. As this effect is not present in the template-matching procedure, the jump no longer appears. 
 
The full corrected radial velocity sets for Gl96 and Gl617A are given in Appendix \ref{A:radvel}, in Tables \ref{Gl96_RV_t}, \ref{Gl617A_RV_t} respectively.
 
\subsection{Activity Indicators}

In order to study the stellar activity of our targets, we calculated several indicators. Two of the most widely used spectral indicators of chromospheric activity are the H$\rm \alpha$ index, which measures the flux in the H$\rm \alpha$ line, and the $\rm log(R'_{HK})$ index, which is based on the flux in the Ca II H and K emission lines. We computed the H$\alpha$ index following the definitions of \cite{Boisse09}. For the Ca lines, we followed \cite{Boisse10} to obtain the S-index scaled to Mount Wilson values. The $\rm log(R'_{HK})$ index was originally defined by \cite{Noyes}, using a photometric correction based on the B-V index. However, in that work the conversion from the S-index to $\rm log(R'_{HK})$ was not calibrated for redder M-dwarfs, and the B-V index is not ideal for M-dwarfs which are too faint in the B band. Therefore, we used the calibrations of \cite{Astudillo17} for M-dwarfs to calculate the $\rm log(R'_{HK})$ index employing V-K colours.

\cite{Gomes11} carried out a study of activity indices for M-dwarfs, finding that - in addition to the H$\rm \alpha$ and $\rm log(R'_{HK})$ indices - the Na I D1 and D2 lines correlate well with stellar activity in these stars. Therefore, we also calculated the NaI index as defined by the authors.

The CCF bisector is also known to correlate with stellar activity for short rotational periods. We obtained the bisector for our observations from the SOPHIE pipeline.

\subsection{Stellar Parameters}
\label{s:starparam}

The stellar parameters are listed in Table \ref{starprop}. Spectral types were obtained from \cite{Gaidos}; masses from \cite{Delfosse}; metallicities and luminosities from \cite{GaidosMann}; temperatures from \cite{Mann15} when available, and if not from \cite{GaidosMann}. Magnitudes and colour indices were taken from \cite{Zacharias}, except for the K magnitude which is from \cite{Cutri}. For Gl617A, a GAIA DR1 parallax is available, while for Gl96 we take the HIPPARCOS parallax. The mean and standard deviation of $\rm log(R'_{HK})$ were calculated from the SOPHIE spectra. We used the $\rm log(R'_{HK})-log(P_{rot})$ relation from \cite{Astudillo17} to estimate the rotation period from the mean $\rm log(R'_{HK})$, with error bars calculated by propagation.

\begin{table}[h]
\caption[]{\label{starprop}Stellar parameters.}
\begin{tabular}{lcc}
\hline \hline
Parameter & Gl617A & Gl96 \\ \hline
Spectral Type	&	M1\tablefootmark{a}	&	M2\tablefootmark{a}	\\
V	&	8.896\tablefootmark{b}	&	9.345\tablefootmark{b}	\\
B-V	&	1.010\tablefootmark{b}	&	1.519\tablefootmark{b}	\\
V-K	&	3.943\tablefootmark{c}	&	3.791\tablefootmark{c}	\\
Mass [$M_\odot$] &	$0.60 \pm 0.07$\tablefootmark{d}	&	$0.60 \pm 0.07$\tablefootmark{d}	\\
$\Pi$ [mas]	&	$93.15 \pm 0.23$\tablefootmark{e}	&	$83.75 \pm 1.14$\tablefootmark{f}	\\
$log(R'_{HK})$	&	$-4.75 \pm 0.14$	&	$-4.77 \pm 0.06$	\\
$P_{rot}$ [d]	&	$28.8 \pm 6.1$	&	$29.6 \pm 2.8$	\\
$T_{eff}$ [k]	&	$4156 \pm 73$\tablefootmark{g}	&	$3785 \pm 62$\tablefootmark{h}	\\
$L_{\star}$ [$L_\odot$]	&	$0.1069 \pm 0.0153$\tablefootmark{g}	&	$0.0888 \pm 0.0135$\tablefootmark{g}		\\
Fe/H [dex]	&	$0.19 \pm 0.08$\tablefootmark{g}	&	$0.14 \pm 0.08$\tablefootmark{g}	\\
\hline
\end{tabular}
\tablefoot{
\tablefoottext{a}{Source: \cite{Gaidos}.}
\tablefoottext{b}{Source: \cite{Zacharias}.}
\tablefoottext{d}{Source: V: \cite{Zacharias}, K: \cite{Cutri}.}
\tablefoottext{d}{Source: \cite{Delfosse}.}
\tablefoottext{e}{Source: \cite{GAIA}.}
\tablefoottext{f}{Source: \cite{VanLeeuwen}.}
\tablefoottext{g}{Source: \cite{GaidosMann}.}
\tablefoottext{h}{Source: \cite{Mann15}.}
}
\end{table}

\subsection{Radial velocity analysis}

To analyse the radial velocities, we employed the DACE (Data and Analysis Center for Exoplanets) web platform\footnote{Available at \url{https://dace.unige.ch}}, which is based at the University of Geneva. This platform allows the generation of generalized Lomb-Scargle periodograms and the calculation of false alarm probabilities (FAP). Signals can be fit by keplerian models (DACE employs the formalism of \citealt{Delisle} for this purpose), and polynomial drifts and stellar jitter can be added. DACE also provides a MCMC analysis facility, described in \cite{Diaz14} and \cite{Diaz16}.

%--------------------------------------------------------------------

\section{Results}
\label{s:res}

\subsection{Gl96}

\begin{figure}
\centering
\includegraphics[width=\hsize]{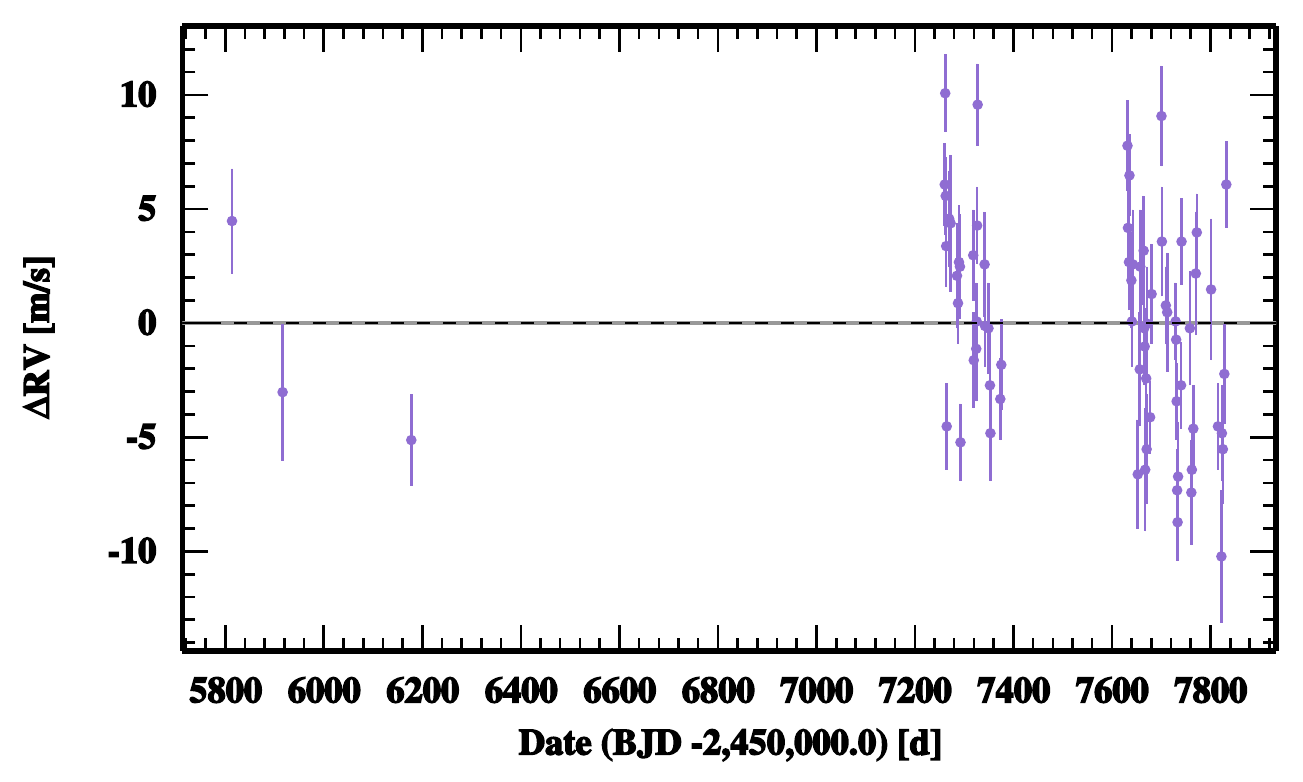}
\caption{Time series of the radial velocities calculated with template-matching for Gl96 from the SOPHIE+ measurements.}
\label{Gl96_series}
\end{figure}

\begin{figure}
\centering
\includegraphics[width=\hsize]{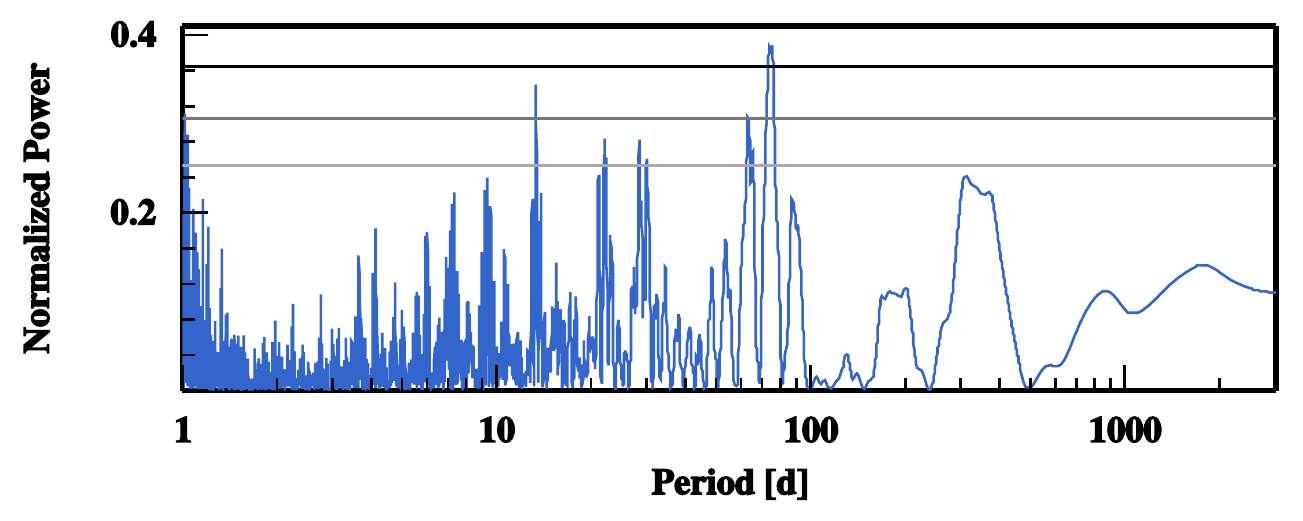}
\includegraphics[width=\hsize]{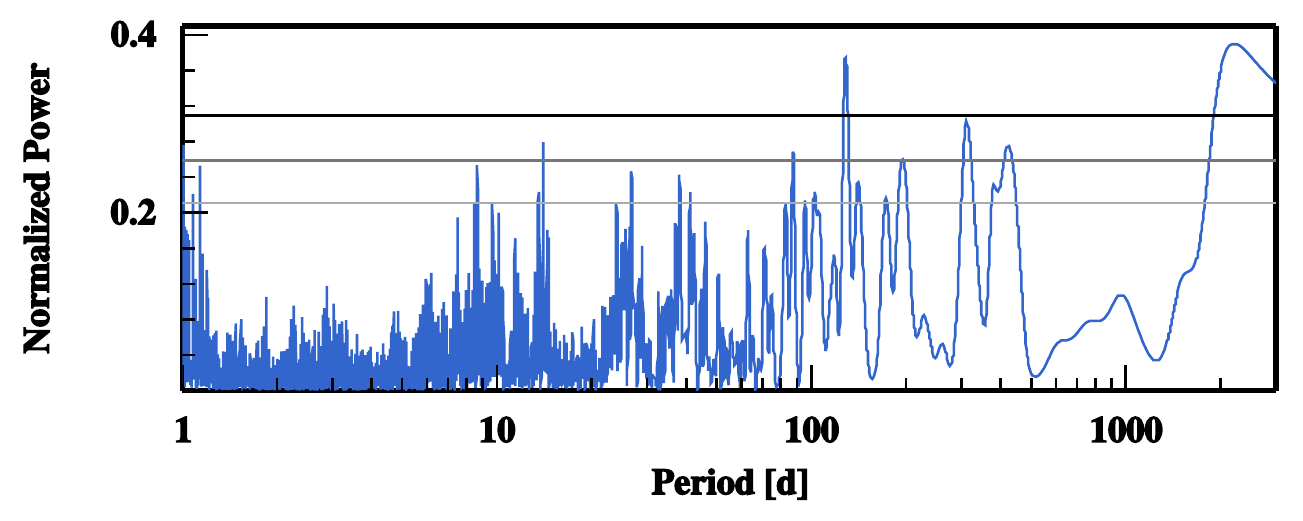}
\includegraphics[width=\hsize]{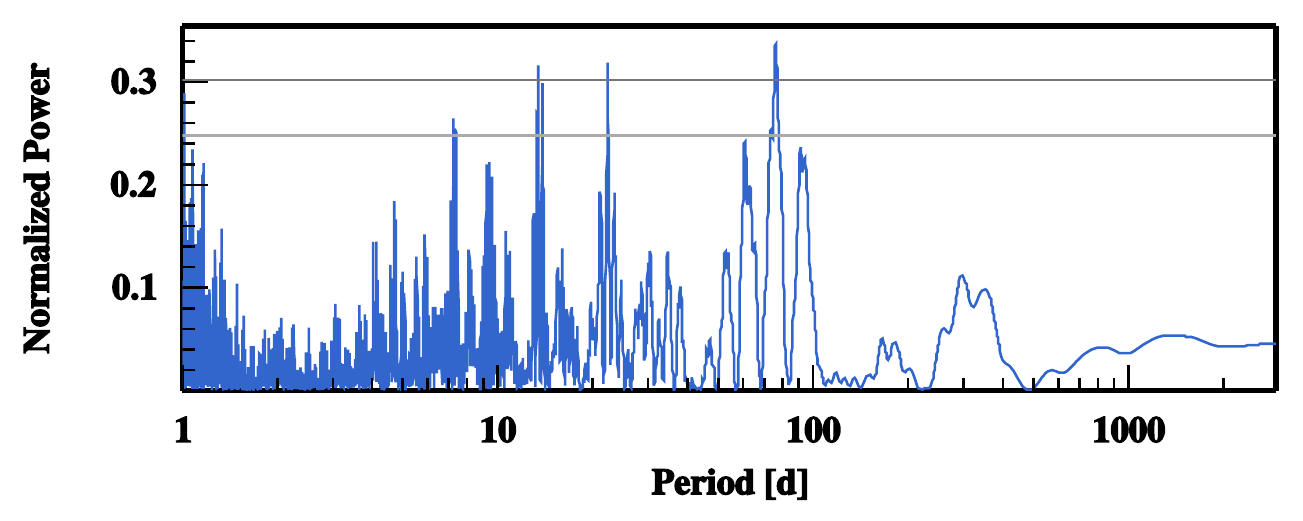}
\caption{Periodogram of (top) the radial velocities calculated with template-matching for Gl96 from the SOPHIE+ measurements, corrected from the zero-point drift; (middle) the time series of the master correction for the zero-point drift applied; (bottom) the uncorrected radial velocities for Gl96 (prior to the application of the master correction). The horizontal lines correspond to 50\%, 10\%, and 1\% FAP respectively.}
\label{Gl96_per}
\end{figure}

\begin{figure}
\centering
  \includegraphics[width=\hsize]{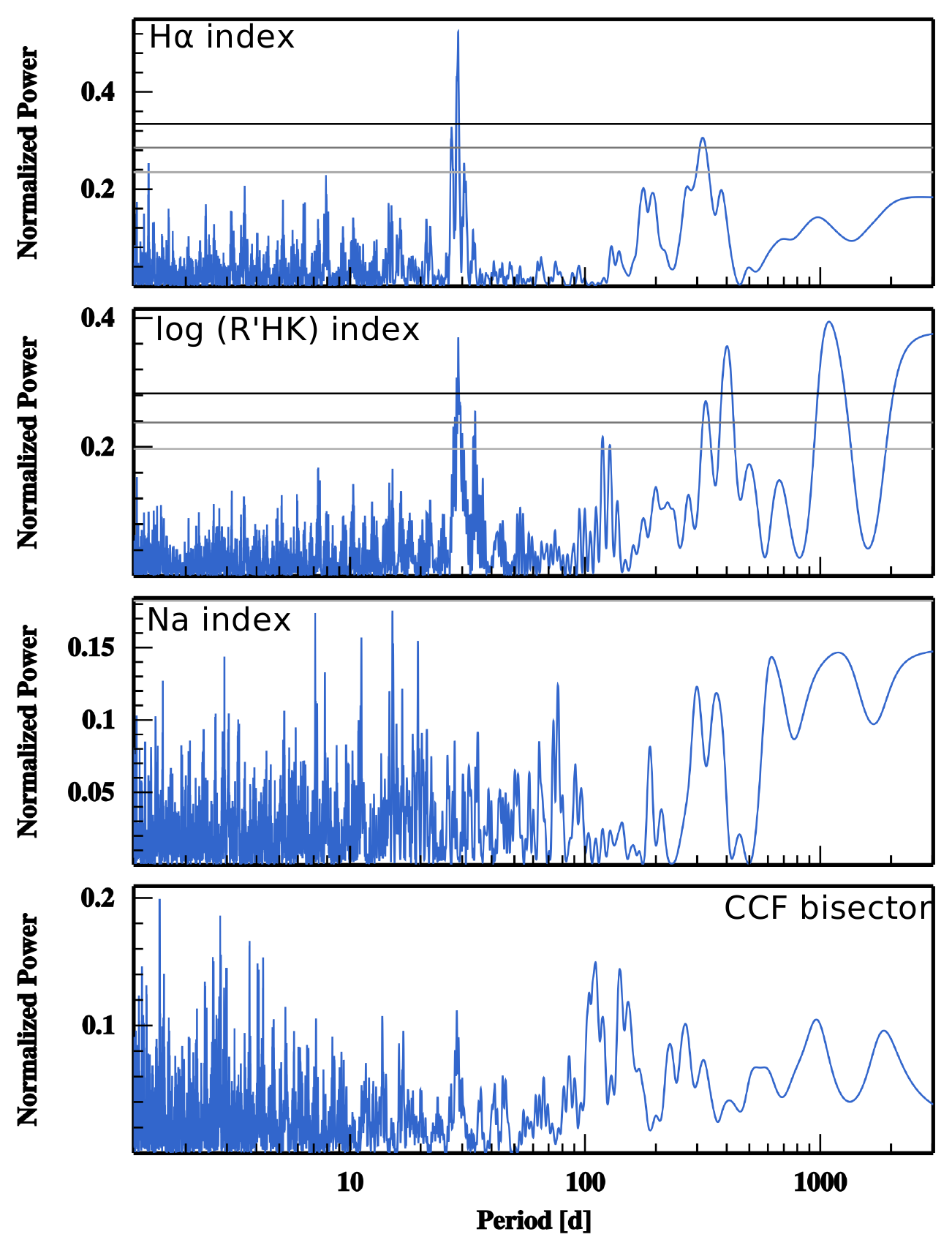}
\caption{Periodograms of activity indicators for Gl96: from top to bottom, H$\alpha$ index, $\rm log(R'_{HK})$ index, NaI index, and CCF bisector. The horizontal lines correspond to 50\%, 10\%, and 1\% FAP respectively. For the NaI index and CCF bisector, no horizontal lines are visible because the entire periodogram is beneath the 50\% FAP line.}
\label{Gl96_act}
\end{figure}

The RVs calculated with template-matching from the SOPHIE+ observations of Gl96 were analysed using DACE. The time series and periodogram of the Gl96 RVs are shown in Figures \ref{Gl96_series} and \ref{Gl96_per} respectively. Figure \ref{Gl96_per} also shows the periodogram of the zero-point correction applied, and of the data prior to this correction. In Figure \ref{Gl96_act} we show periodograms of the activity indicators H$\alpha$, $\rm log(R'_{HK})$, NaI, and the CCF bisector.

\begin{figure}
\centering
\includegraphics[width=\hsize]{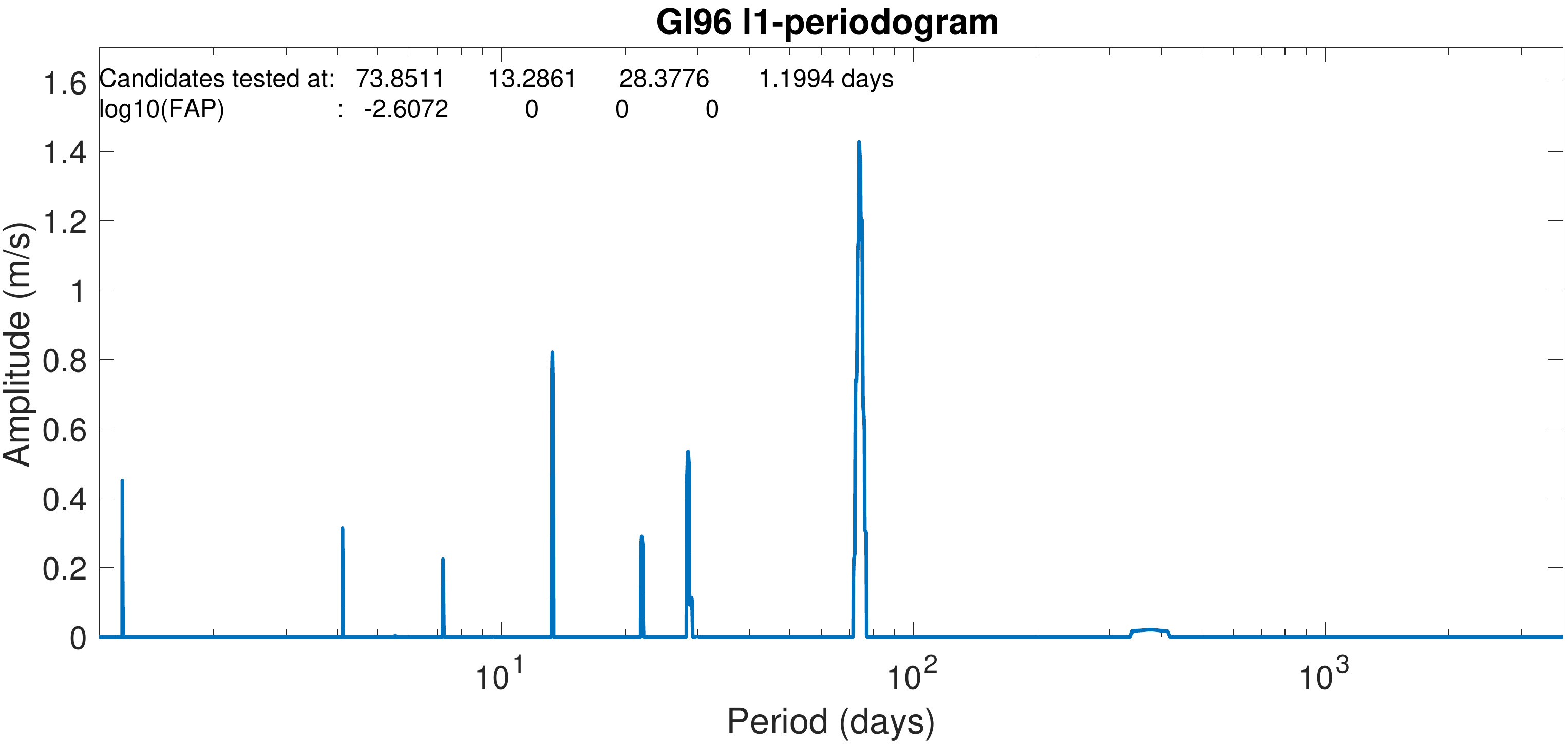}
\caption{l1 periodogram of the SOPHIE RVs for Gl96. The signal at 74 days is clearly predominant, while the other two peaks are probably related to activity.}
\label{Gl96-l1}
\end{figure}

The RV periodogram shows a peak at 75d below 1\% FAP, which bootstrap resampling places below 0.05\% FAP. There is no corresponding peak for any of the activity indicators, nor does the zero-point drift correction applied show any signal at this period. Additionally, the signal also appears in the uncorrected time series periodogram. We also applied an l1 periodogram, as defined by \cite{Hara}; this technique searches for a number of signals simultaneously, using compressed sensing techniques. As such, it is much less prone to aliases and other problems arising in the traditional periodogram. The resulting periodogram is shown in Figure \ref{Gl96-l1}, where the signal at 74 days can clearly be seen to dominate the data; the FAP of this signal is conservatively estimated (using an analytical formula from \citealt{Baluev08}) as $log_{10}(FAP) = -2.6072$. The remaining signals are consistent with the stellar rotation period and half this period, suggesting they originate in stellar activity.

\begin{figure}
\centering
\includegraphics[width=\hsize]{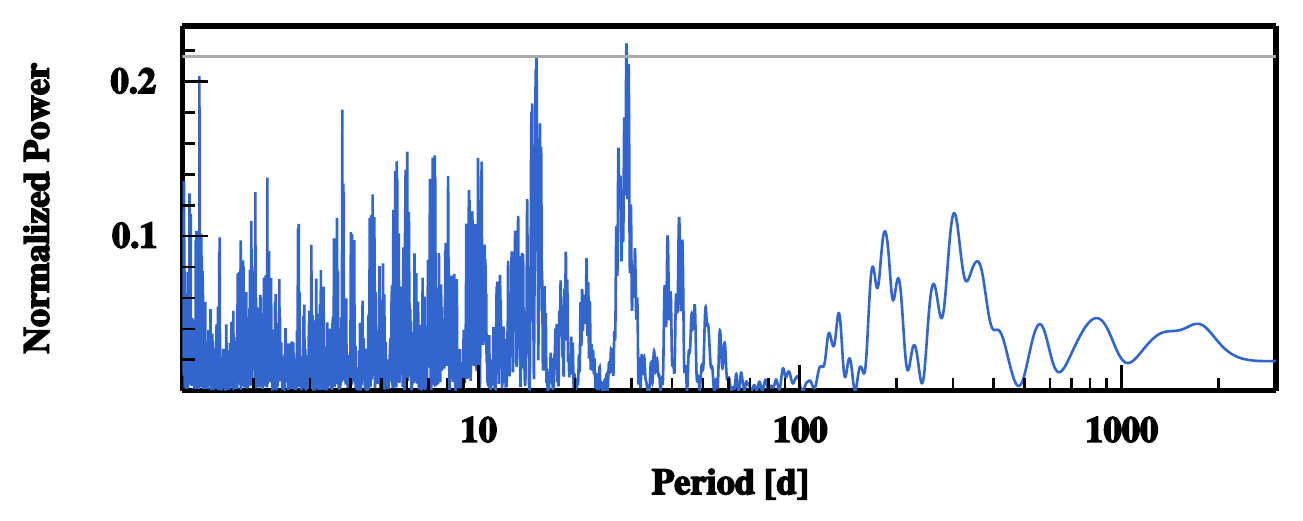}
\caption{Periodogram of the residuals of a keplerian fit with P=75d to the radial velocities calculated with template-matching for Gl96 from the SOPHIE+ measurements. The horizontal line corresponds to 50\% FAP level.}
\label{Gl96_k1_res}
\end{figure}

We fit this signal by a keplerian model with DACE. The highest peak in the residuals, at 29d, is only below 50\% FAP, while a second peak at 14d at similar FAP corresponds to half this period (Figure \ref{Gl96_k1_res}). Additionally, the highest peak is close to the peaks seen at 28-29d well below 1\% FAP in the H$\rm \alpha$ and $\rm log(R'_{HK})$ indices' periodograms. Furthermore, these periods are consistent with the estimated stellar rotation period of $29.6 \pm 2.8$d (see Section \ref{s:starparam}). Therefore, we cannot justify treating it as a potential second planet.

\begin{figure}
    \centering
    \includegraphics[width=\hsize]{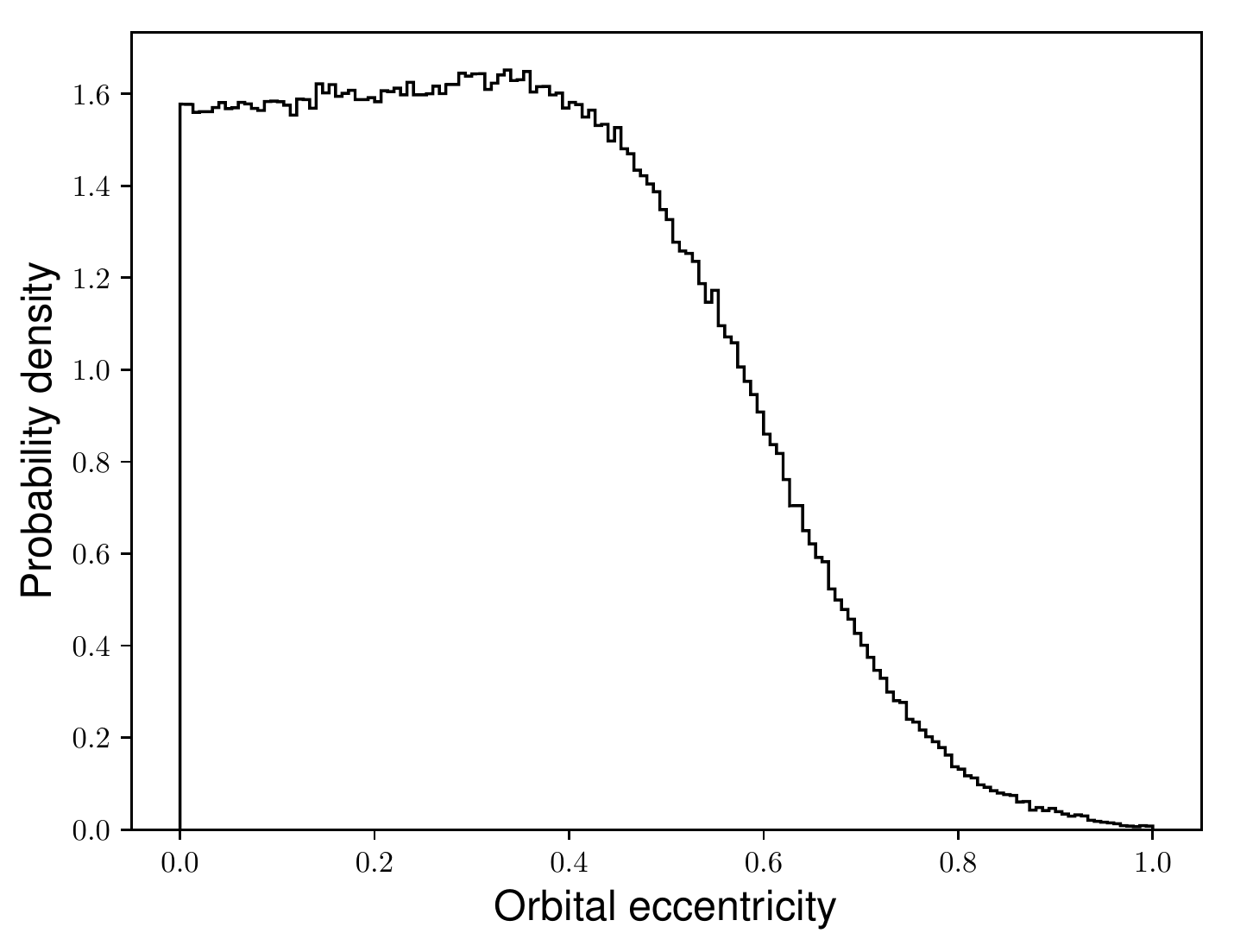}
    \caption{Posterior probability of the orbital eccentricity of Gl96 b under the model including correlated noise. }
    \label{fig:histecc96}
\end{figure}

\begin{figure}
\centering
\includegraphics[width=\hsize]{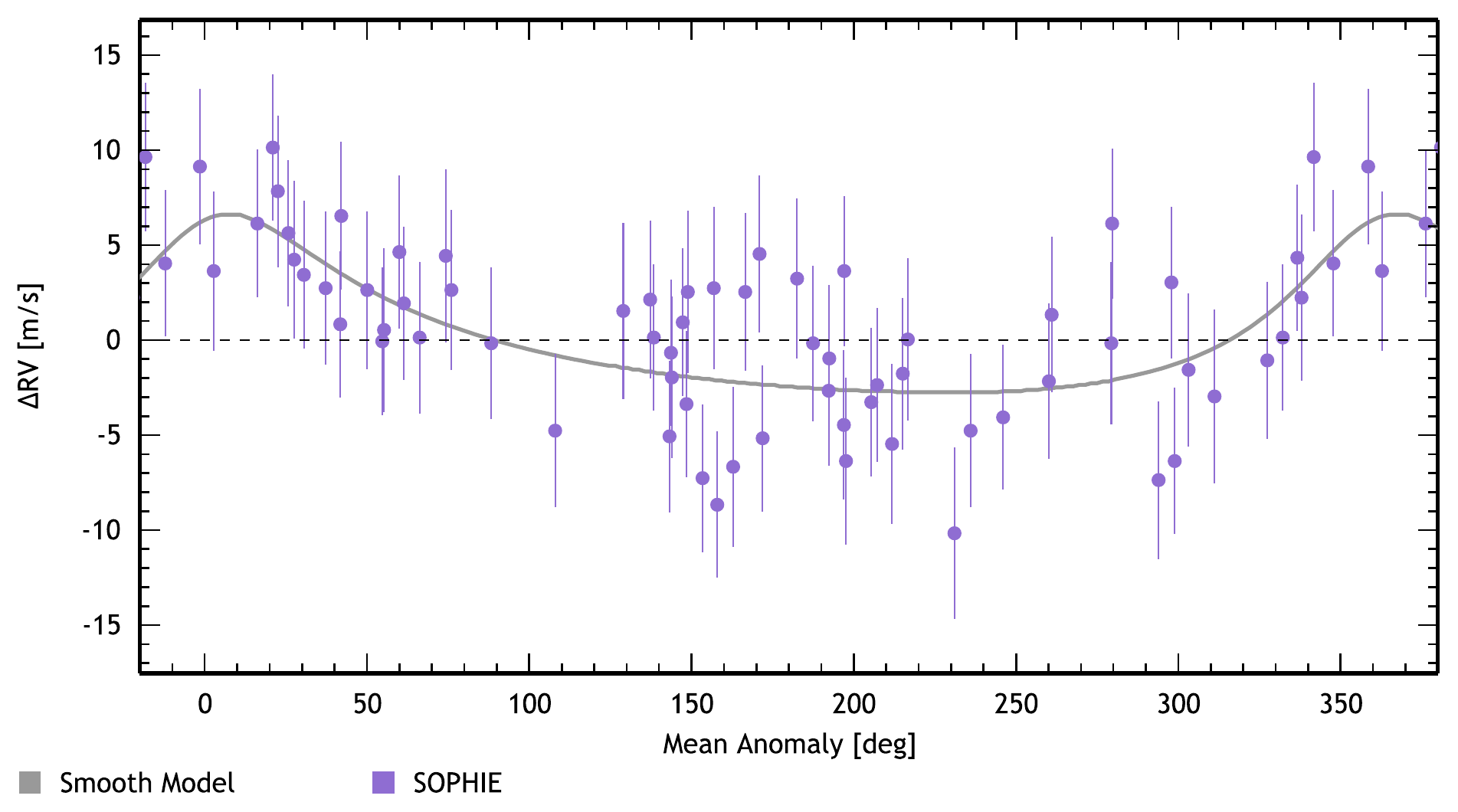}
\caption{Phase-folded radial velocities of Gl96 for a one-planet model with P=74.7d.}
\label{Gl96_k1_phase}
\end{figure}

We used DACE to carry out an MCMC analysis of the single-planet model for Gl96, in order to better constrain the parameters. The results are summarized in Table \ref{Gl96_tab-mcmc-Summary_params}. Figure \ref{Gl96_k1_phase} shows the phase-folded data points. The best-fit solution results in a rather highly eccentric orbit. In order to analyse whether contamination from stellar-activity driven RV variations is influencing the results, we tested two approaches: the addition of a keplerian fit to the 29d peak, and a red-noise model. The two-keplerian model did not modify the planetary parameters greatly; in particular, the resulting eccentricity is of 0.46$_{-0.14}^{+0.22}$, which is indistinguishable from the one-keplerian result within error bars. The red noise was modelled using a Gaussian process with a quasi-periodic kernel \citep[see details of the model in, e.g.,][]{Astudillo17c}. We informed the model of the rotational period of the star by including an informative prior on the corresponding hyper-parameter. The posterior distribution of this parameter is narrower, and we find $P_{rot} = 28.4\pm1.4$ days, where the uncertainties correspond to the 1-$\sigma$ credible interval. Concerning the planet eccentricity, the maximum-a-posteriori estimate is 0.50, in agreement with the model without red noise. The inclusion of correlated noise, on the other hand, seems to allow for lower values of the eccentricity: the 95-\% highest-density interval extends between 0.0 and 0.68 (Fig.~\ref{fig:histecc96}), showing that in the presence of red noise, the eccentricity is effectively unconstrained. 

\begin{table}
\caption{Best-fitted solution for the planetary system orbiting Gl96.}  \label{Gl96_tab-mcmc-Summary_params}
\begin{tabular}{lcc}
\hline
\hline
Param. & Units & Gl96 b \\
\hline \\[-1.5ex]
$P$ & [d] & 73.94$_{-0.38}^{+0.33}$  \\ [0.5ex]
$K$ & [m\,s$^{-1}$] & 4.69$_{-0.62}^{+0.72}$  \\[0.5ex]
$e$ &   & 0.44$_{-0.11}^{+0.09}$  \\[0.5ex]
$\omega$ & [deg] & 339.58$_{-14.52}^{+12.45}$  \\[0.5ex]
$T_P$ & [d] & 55556.39$_{-8.98}^{+10.57}$  \\[0.5ex]
$T_C$ & [d] & 55568.90$_{-9.82}^{+11.56}$  \\[0.5ex]
\hline \\[-1.5ex]
$Ar$ & [AU] & 0.291$_{-0.005}^{+0.005}$  \\[0.5ex]
%M.$\sin{i}$ & [M$_{\rm Jup}$] & 0.0619$_{-0.0072}^{+0.0076}$  \\
M.$\sin{i}$ & [M$_{\rm Earth}$] & 19.66$_{-2.30}^{+2.42}$  \\[0.5ex]
\hline \\[-1.5ex]
$\gamma_{SOPHIE}$ & [m\,s$^{-1}$] & \multicolumn{1}{c}{-37874.84$_{-0.32}^{+0.31}$}\\[0.5ex]
$\sigma_{JIT}$ & [m\,s$^{-1}$] & \multicolumn{1}{c}{3.45$_{-0.91}^{+0.93}$}\\[0.5ex]
$\sigma_{(O-C)}$ & [m\,s$^{-1}$] & \multicolumn{1}{c}{3.37}\\[0.5ex]
$\log{(\rm Post})$ &   & \multicolumn{1}{c}{-196.77$_{-2.46}^{+1.68}$}\\[0.5ex]
\hline
\end{tabular}
\tablefoot{ For each parameter, the median of the posterior is reported, with error bars computed from the MCMC chains using a 68.3\% confidence interval. $\sigma_{O-C}$ corresponds to the weighted standard deviation of the residuals around this best solutions. $\log{(\rm Post})$ is the posterior likelihood. All the parameters probed by the MCMC can be found in Appendix \ref{a:MCMC}, Table \ref{Gl96_tab-mcmc-Probed_params}}
\end{table}

\begin{figure}[htb]
\centering
\includegraphics[height=0.41\hsize]{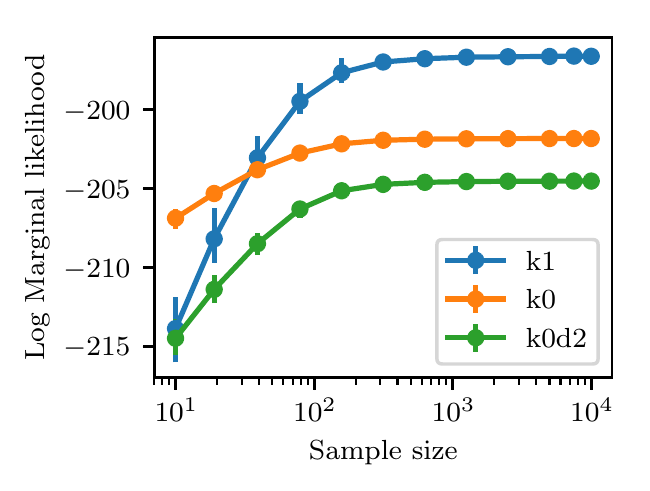}
\includegraphics[height=0.41\hsize]{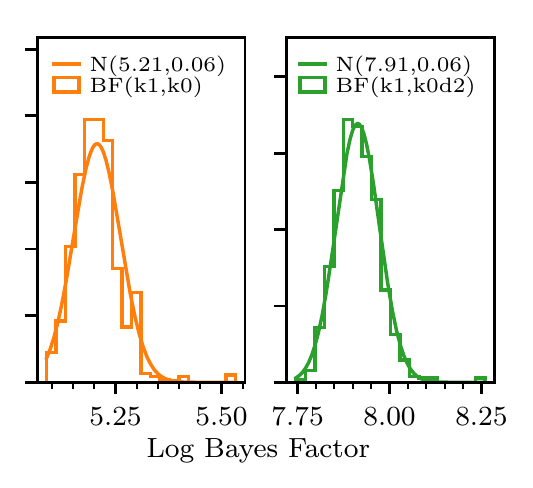}
\caption{\emph{Left}: Marginal likelihood, estimated by the Perrakis method, vs. sample size for three models: a single Keplerian plus white noise (k1; blue), a quadratic drift plus white noise (k0d2; green), and pure white noise (k0; orange). The one-Keplerian model is clearly favoured by the data. The error bars correspond to the 95\% confidence interval. \emph{Right}: histogram of 1000 MonteCarlo realisations of the Bayes Factor between model k1 and k0 (orange) and k0d2 (green); the solid curves are normal distributions with the mean and variance equal to those of the MonteCarlo sample.}
\label{Gl96-ML}
\end{figure}

To quantify the significance of the detection, we estimated the posterior odds ratio (POR) between several competing models. If we assume equal prior probability for all models, the POR reduces to the ratio of marginal likelihoods, i.e. the Bayes factor ($BF$). We used the estimation introduced by \citet{Perrakis2014}, based on importance sampling, to compute the marginal likelihoods for a model without any Keplerian (k0), a model with a quadratic long-term trend (k0d2), and a model with a single Keplerian (k1). All three models included an additional white noise component, whose amplitude was an additional nuisance parameter. The results are shown in Figure \ref{Gl96-ML} as a function of the sample size used for the estimation. As expected, the Perrakis estimator is biased, but for sample sizes larger than around 3000, the bias is negligible. We find logarithmic Bayes factors $\log(BF_\mathrm{k1; k0}) = 5.21 \pm 0.06$, and $\log(BF_\mathrm{k1; k0d2}) = 7.91 \pm 0.06$, for the comparison between k1 and k0, and k1 and k0d2, respectively, where the reported values are the empirical means and standard deviations obtained by repeating the calculation for each model 5000 times, and drawing 1000 random pairs. The resulting distributions are very approximately normal. The analysis shows, therefore, that the posterior probability of a model with a single Keplerian is much larger than any of the competing models. This is so in spite of the strong penalisation the Bayes factor gives to more complex models.

\begin{figure}
\centering
\includegraphics[width=\hsize]{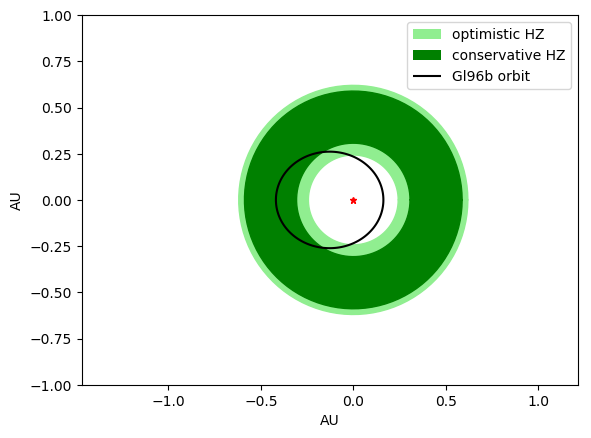}
\caption{Calculated orbit for Gl96 b with respect to the optimistic and conservative habitable zones, as defined by \cite{Kopparapu13}.}
\label{Gl96_HZ}
\end{figure}

One of the goals of the SP3 programme is to detect habitable planets around M-dwarf stars. The habitable zone calculator\footnote{Located at \url{https://depts.washington.edu/naivpl/content/hz-calculator}} based on the work of \cite{Kopparapu13} places Gl96 b inward of the conservative HZ, though within the optimistic one; however, due to its high eccentricity it would probably move inward of the optimistic HZ at periapsis (Figure \ref{Gl96_HZ}). To better quantify the habitability of this eccentric planet, we follow the method employed by \cite{Diaz16b}, who calculated the mean incident flux over an orbit as defined by \cite{Williams}, and compared it with the limits given by \cite{Kopparapu13b}. For Gl96 b, the mean incident flux is $<F>/F_ \oplus = 1.168$, placing it between the Recent Venus and Runaway Greenhouse limits.

\subsection{Gl617A}

A planet at 86.54 d around this star was recently announced by the CARMENES team \citep{Reiners}. In this section, we present an independent detection of this planet from our SOPHIE data, and describe a further potential candidate at 500 d.

\begin{figure}
\centering
\includegraphics[width=\hsize]{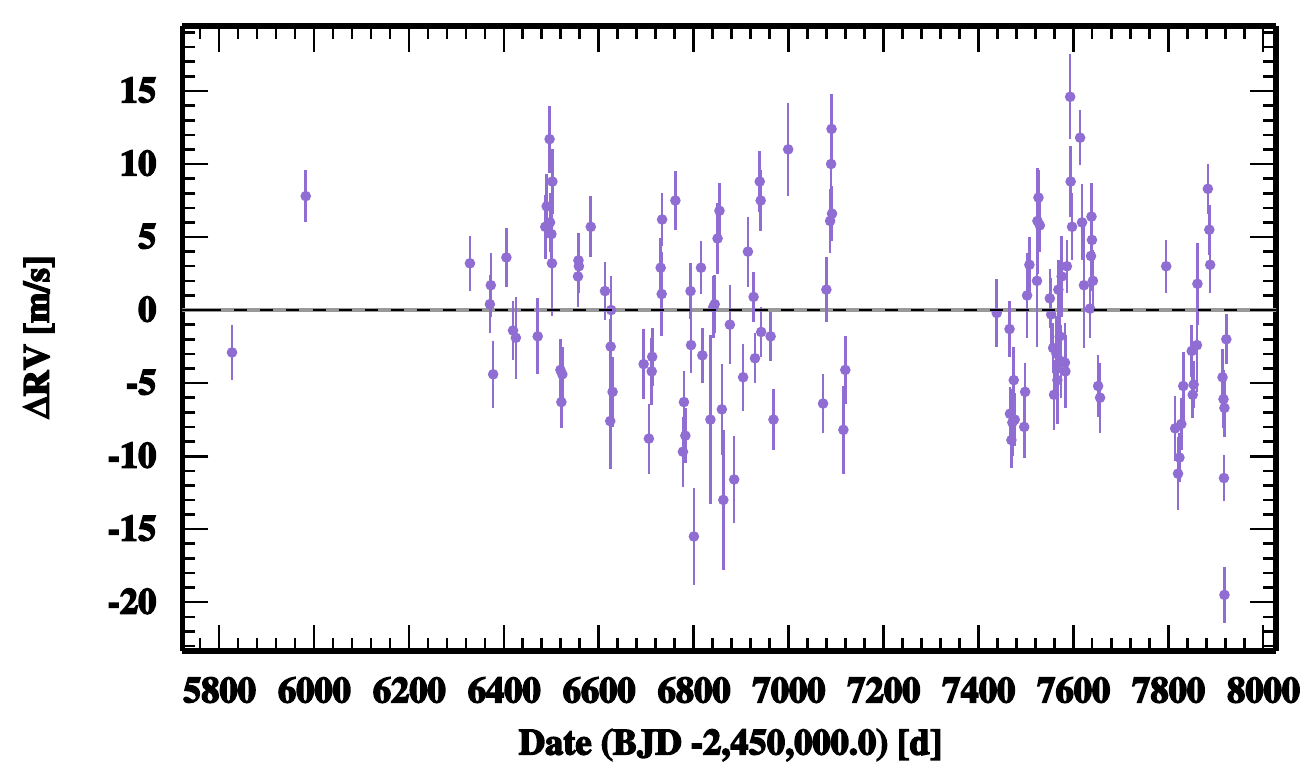}
\caption{Time series of the radial velocities calculated with template-matching for Gl617A from the SOPHIE+ measurements.}
\label{Gl617A_series}
\end{figure}

\begin{figure}
\centering
\includegraphics[width=\hsize]{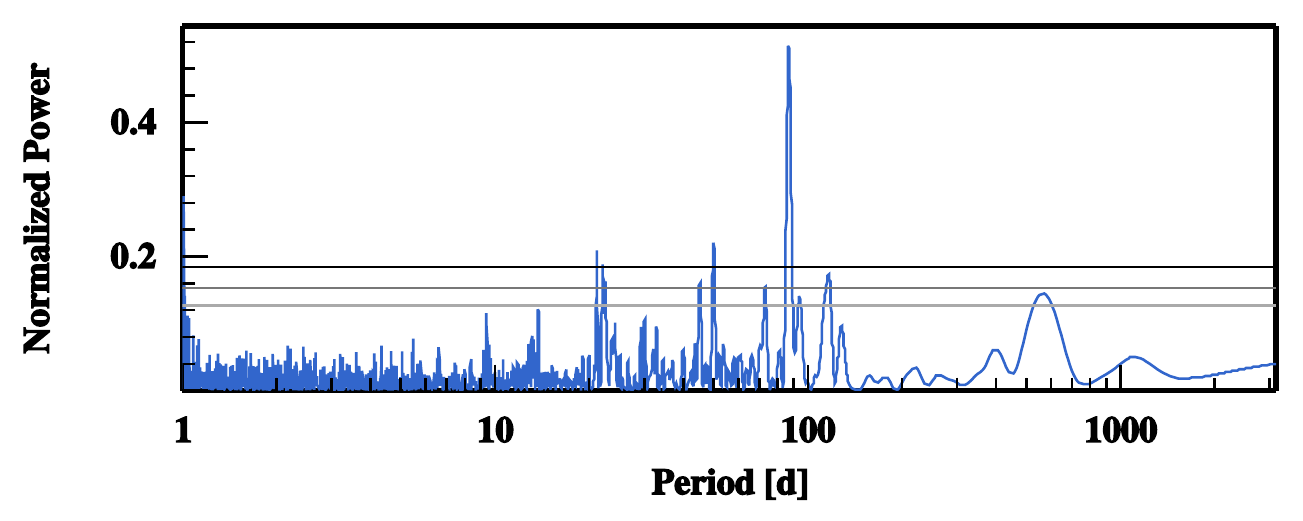}
\includegraphics[width=\hsize]{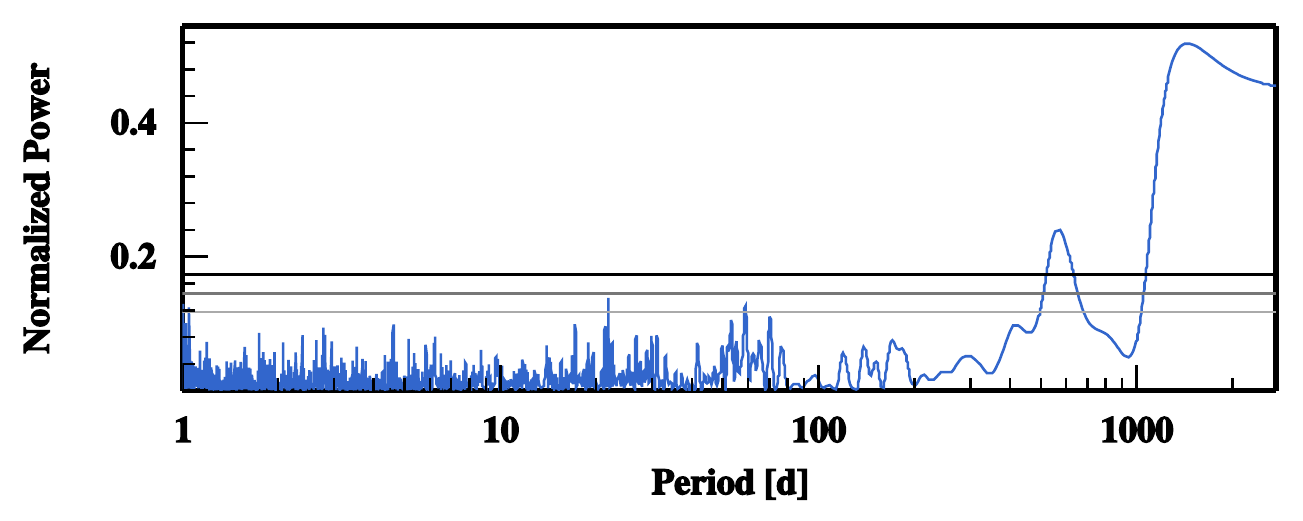}
\includegraphics[width=\hsize]{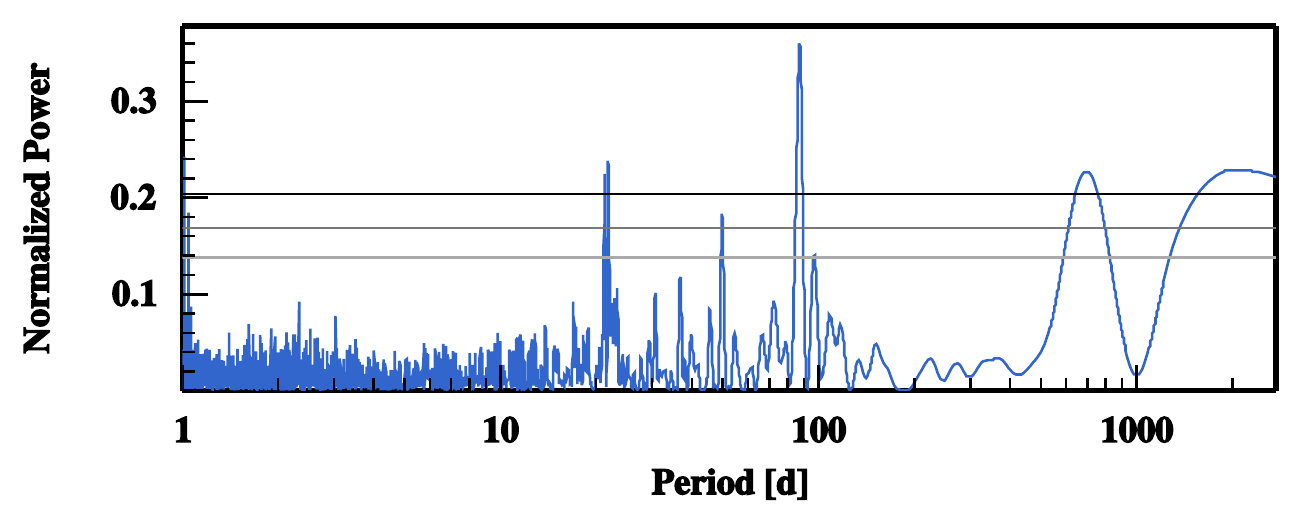}
\caption{Periodogram of: (top) the radial velocities calculated with template-matching for Gl617A from the SOPHIE+ measurements, corrected from the zero-point drift; (middle) the master correction for the zero-point drift applied; (bottom) the uncorrected radial velocities for Gl617A (prior to the application of the master correction). The horizontal lines correspond to 50\%, 10\%, and 1\% FAP respectively.}
\label{Gl617A_per}
\end{figure}

We analysed the RVs calculated by template-matching from the SOPHIE+ observations of Gl617A with the DACE platform. The time series and periodogram of the Gl617A RVs are shown in Figures \ref{Gl617A_series} and \ref{Gl617A_per} respectively. Figure \ref{Gl617A_per} also shows the periodograms of the zero-point correction applied and of the uncorrected data (i.e. the Gl617A RVs prior to the application of this correction).

\begin{figure}
\centering
 \includegraphics[width=\hsize]{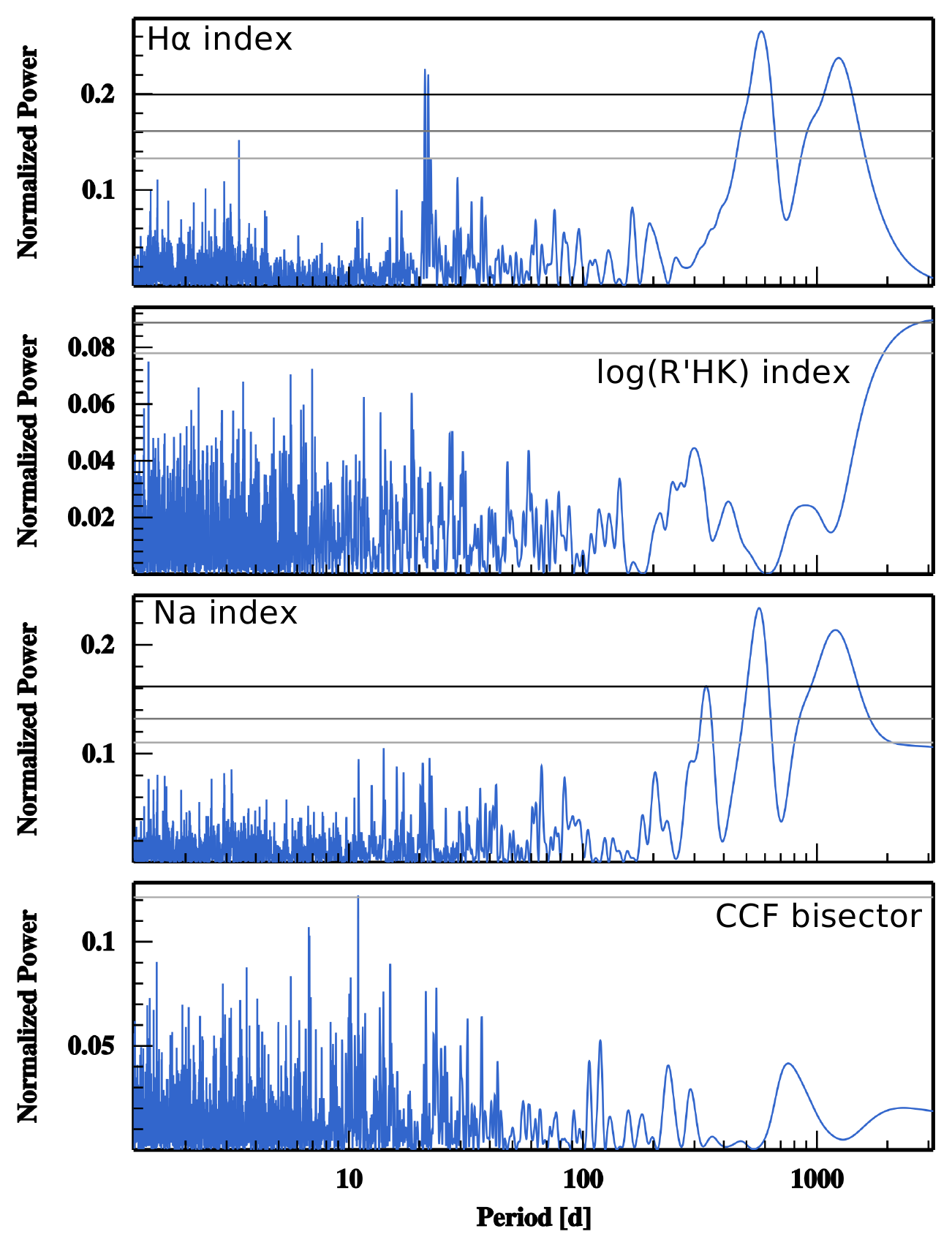}
\caption{Periodograms of activity indicators for Gl617A: from top to bottom, H$\rm \alpha$ index, $\rm log(R'_{HK})$ index, NaI index, and CCF bisector. The horizontal lines correspond to 50\%, 10\%, and 1\% FAP respectively.}
\label{Gl617A_act}
\end{figure}

In Figure \ref{Gl617A_act} we show periodograms of the activity indicators H$\rm \alpha$, $\rm log(R'_{HK})$, NaI, and the CCF bisector. The H$\rm \alpha$ index exhibits a periodicity at around 21.8 days, probably related to the rotational period of the star, as well as long-period peaks. The $\rm log(R'_{HK})$ and NaI periodograms present only very long-period peaks, in the 500-1000 day range. The CCF bisector shows a small peak at 10.9 days, which is close to half the rotation period and therefore is probably due to stellar activity.

\begin{figure}
\centering
\includegraphics[width=\hsize]{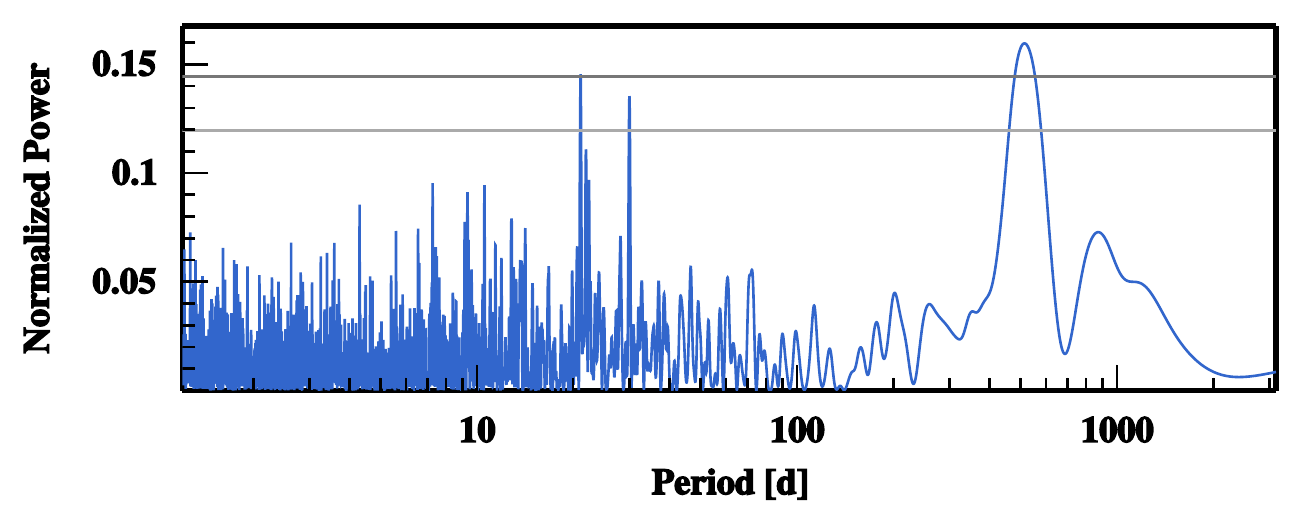}
\caption{Periodogram of the residuals of a keplerian fit with P=86d to the radial velocities calculated with template-matching for Gl617A from the SOPHIE+ measurements. The horizontal lines correspond to 50\% and 10\% FAP respectively.}
\label{Gl617A_k1_res}
\end{figure}

The RV periodogram shows a very strong signal at 86d well below 1\% FAP, which bootstrap resampling places below 0.01\% FAP, and which is not present in any of the activity indicators, nor in the periodogram of the zero-point correction applied. We employed DACE to fit it with a keplerian model. Figure \ref{Gl617A_k1_res} shows the periodogram of the residuals. There are two signals below 10\% FAP: one at 21d and one at around 500d, and one further signal below 50 \% FAP at 29d. The 21d signal is probably due to stellar activity, as it coincides with a signal below 1\% FAP in the H$\rm \alpha$ periodogram, and is close to the estimated rotation period (Table \ref{starprop}), although there is no signal at 21d in the other activity indicators. The 29d signal is removed by the addition of a quadratic drift, suggesting it may be an artifact of the window function (although the period may also point to moon contamination; as this star is observed with simultaneous wavelength calibration, we do not have a measure of the sky). 

The long-period signal is intriguing; the addition of a linear or quadratic drift does not affect it, whereas a test keplerian model removes it completely and results in a low-eccentricity keplerian fit. While the H$\rm \alpha$, $\rm log(R'_{HK})$, and NaI indices' periodograms show long-period peaks, these are not well fitted by the model obtained from the RVs. More concerning is the periodogram of the master correction, which shows a peak at ~500 days; nevertheless, we note that the long-period signal is also present in the uncorrected RVs (see Fig. \ref{Gl617A_per}). However, we stress that this signal is of only moderate significance. We therefore present it as a tentative detection to be confirmed, not a definite planet.

\begin{figure}
\centering
\includegraphics[width=\hsize]{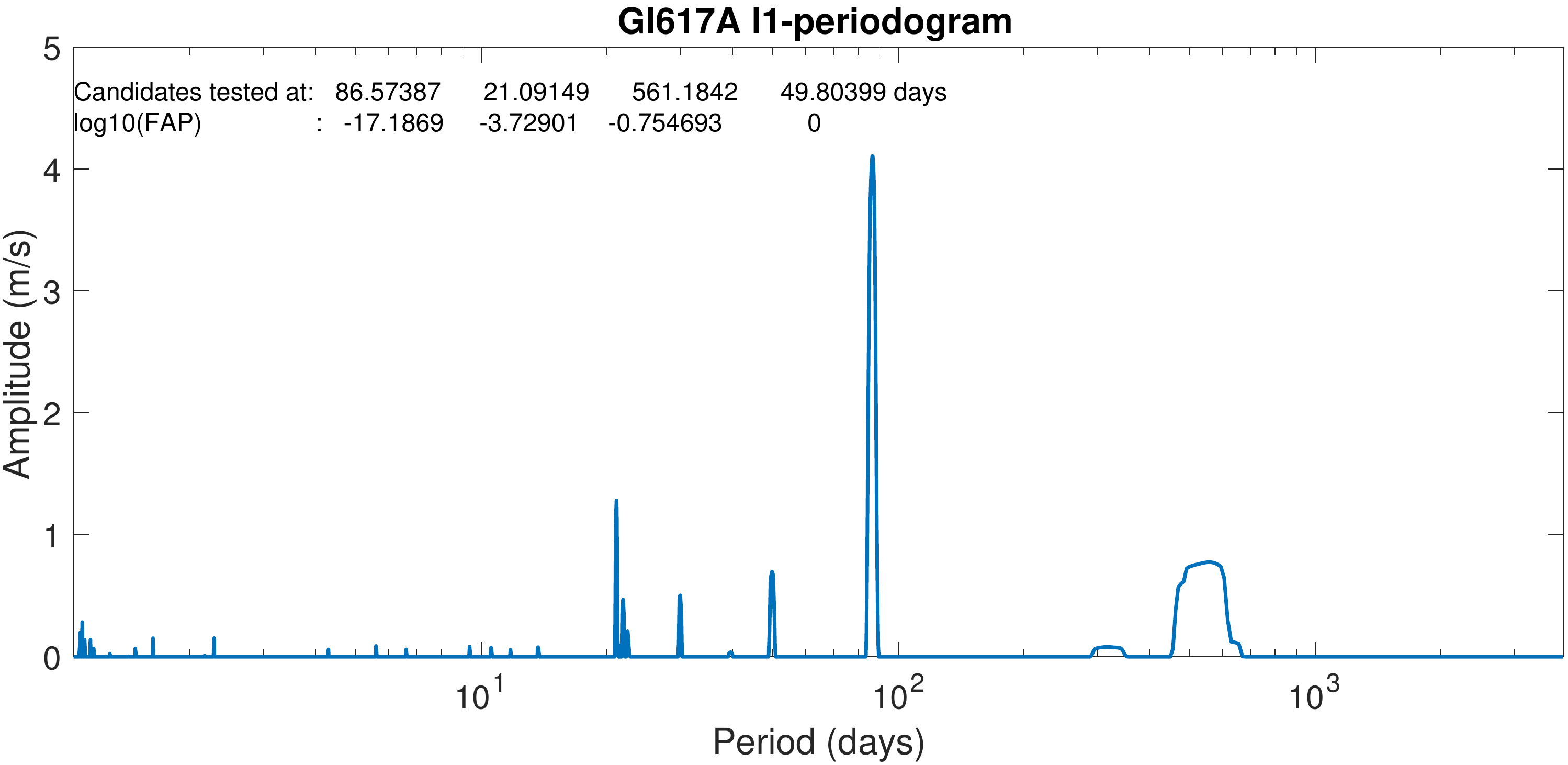}
\caption{l1 periodogram of the SOPHIE RVs for Gl617A. The signal at 86 days is clearly predominant, while the one at 21 days is probably related to activity. The $\sim$500d signal is discussed in the text.}
\label{Gl617A-l1}
\end{figure}

As for Gl96, we also applied an l1 periodogram, which is shown in Figure \ref{Gl617A-l1}; the signal at 86 days dominates the data, with $log_{10}(FAP) = -17.1869$. The remaining signals at 21d and $\sim$500d are consistent with the discussion above.

In order to explore the parameter space more thoroughly, we employed the MCMC sampler implemented in DACE. We tested two models: a single keplerian at 86d, and a two-keplerian model with initial periods at 86d and 500d, both including a quadratic drift. The resulting best-fit solutions are summarised in Table \ref{Gl617A_tab-mcmc-Summary_params_test}.

\begin{figure}
\centering
\includegraphics[width=\hsize]{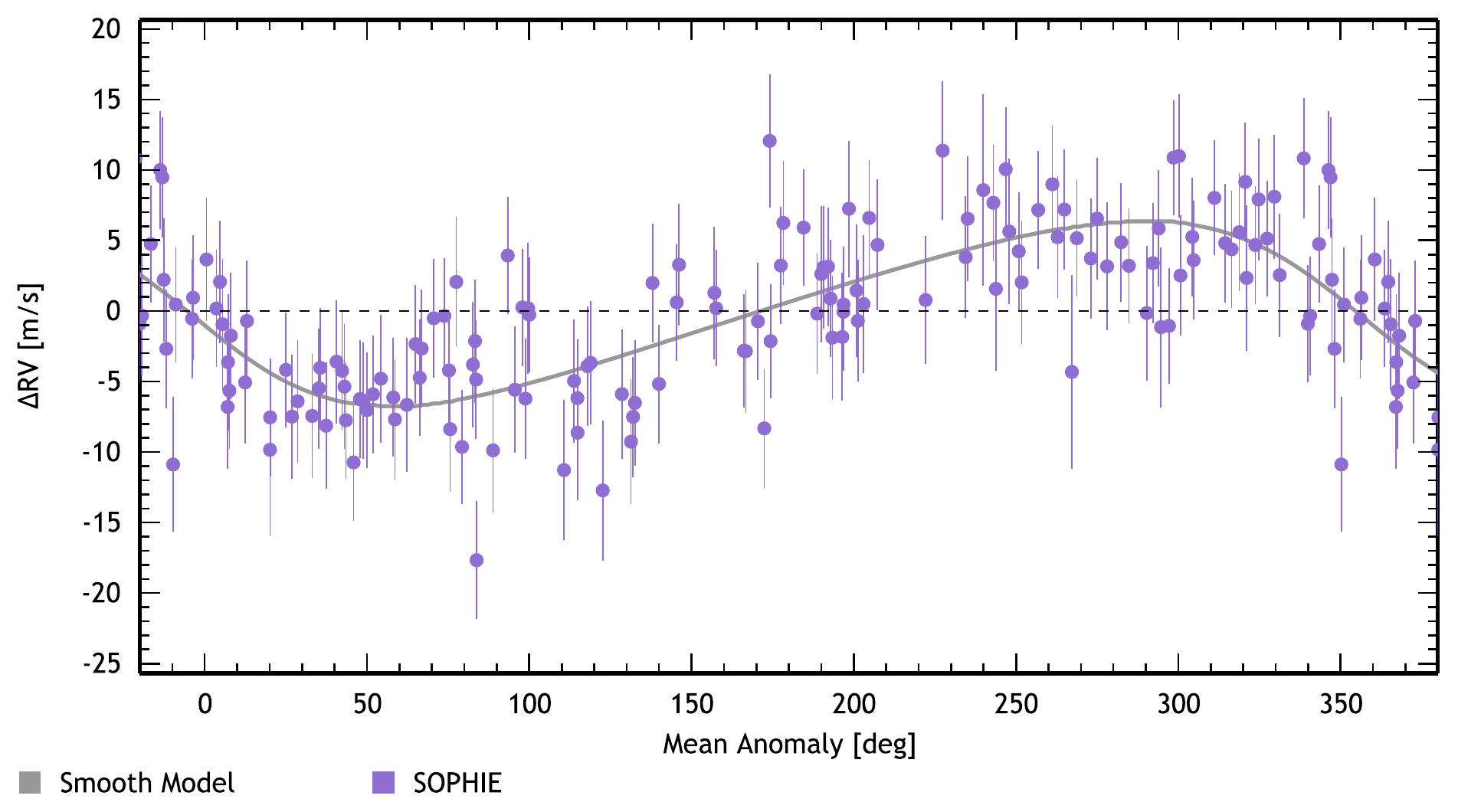}
\caption{Phase-folded radial velocities of Gl617A for a P=86d planet, using the parameters derived from the 2-keplerian model.}
\label{Gl617A_k2_phase}
\end{figure}

\begin{figure}
\centering
\includegraphics[width=\hsize]{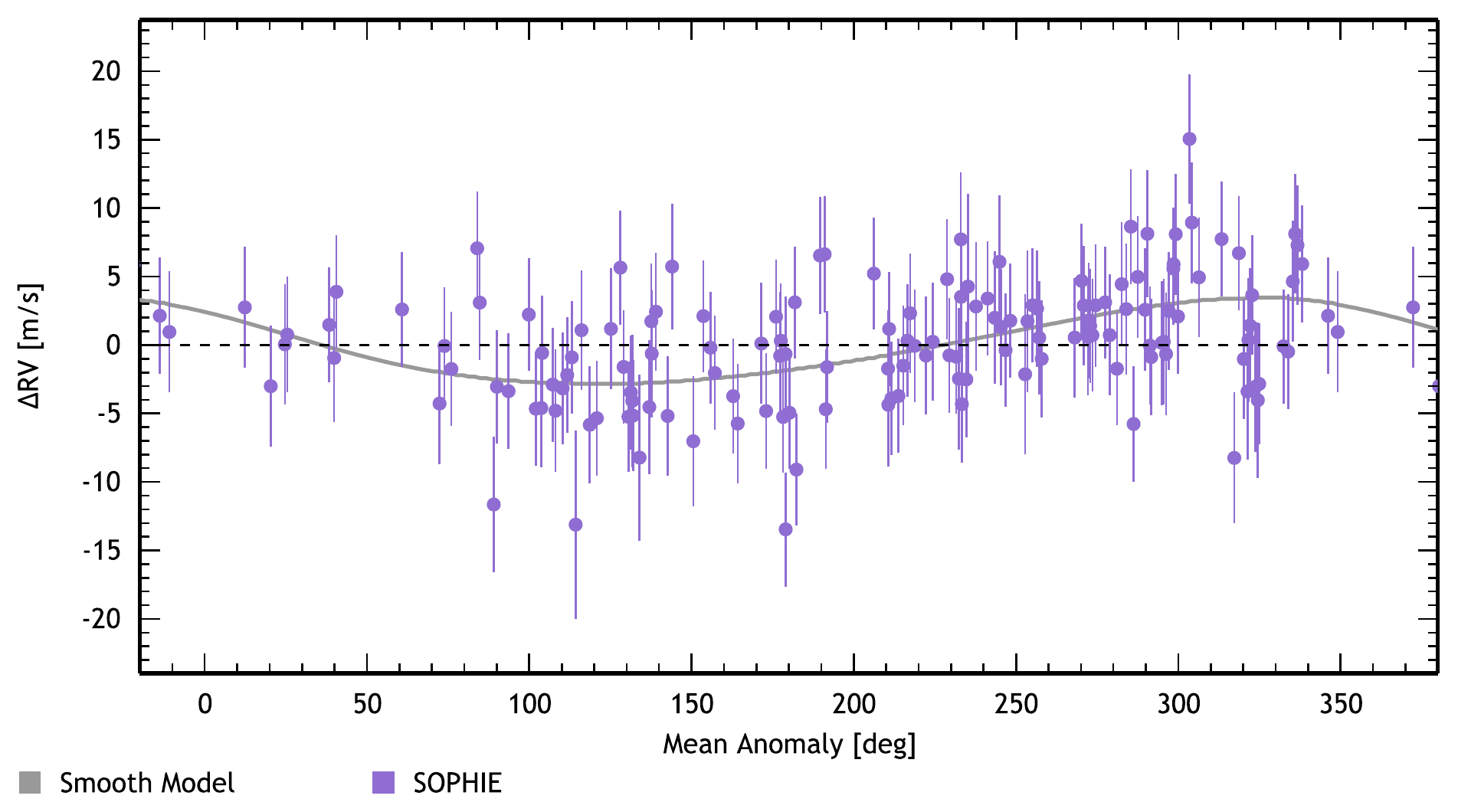}
\caption{Phase-folded radial velocities of Gl617A for a P=497d planet, using the parameters derived from the 2-keplerian model.}
\label{Gl617Ac_phase}
\end{figure}

The planet at 86d is consistent in both models, with practically all parameters indistinguishable within error bars between the two best-fitted solutions. The two-keplerian model actually provides a better fit to the data, with a lower $\rm \sigma_{(O-C)}$ and BIC. Figures \ref{Gl617A_k2_phase} and \ref{Gl617Ac_phase} show the phase-folded RVs for Gl617A b and Gl617A c respectively, using the parameters derived from the two-keplerian model. In the one-keplerian model, the drift is significant; removing it slightly decreases the amplitude and increases the eccentricity of the keplerian fit. In the two-keplerian model, however, the drift is only marginally significant. This is coherent with the fact that in the one-keplerian model, the drift may attempt to absorb the long-period signal that is fitted by a keplerian in the second model.

We compared the parameters obtained by our analysis with those recently presented by \cite{Reiners}. The orbital period and semi-major axis are compatible within error bars; however, we find a somewhat larger mass and distinctly larger eccentricity from our data. The mass reported by \cite{Reiners} of M.$\sin{i}$ = 24.7$_{-1.8}^{+2.4}$ M$_{\rm Earth}$ is compatible with our calculated mass at 2$\sigma$. This is true for both the one-keplerian and the two-keplerian models, with the eccentricity higher for the one-keplerian model (see Table \ref{Gl617A_tab-mcmc-Summary_params_test}).

\begin{table*}
\caption{Best-fitted solutions for the planetary system orbiting Gl617A - 1-keplerian and 2-keplerian models plus drift.}  \label{Gl617A_tab-mcmc-Summary_params_test}
\begin{tabular}{lcccc}
\hline
\hline
Param. & Units & Gl617A b\tablefootmark{1} & Gl617A b\tablefootmark{2} & Gl617A c \\
\hline\\[-1.5ex]
$P$ & [d] & 86.93$_{-0.19}^{+0.18}$ & 86.72$_{-0.18}^{+0.20}$  & 496.90$_{-21.82}^{+35.45}$  \\[0.5ex]
$K$ & [m\,s$^{-1}$] & 6.56$_{-0.39}^{+0.41}$ & 6.57$_{-0.38}^{+0.36}$  & 3.16$_{-0.42}^{+0.43}$  \\[0.5ex]
$e$ &   & 0.32$_{-0.09}^{+0.08}$ & 0.23$_{-0.08}^{+0.07}$  & 0.15$_{-0.10}^{+0.15}$  \\[0.5ex]
$\omega$ & [deg] & 102.20$_{-11.48}^{+10.04}$ & 97.25$_{-13.46}^{+13.55}$  & -311.97$_{-63.26}^{+980.42}$  \\[0.5ex]
$T_P$ & [d] & 55466.60$_{-4.32}^{+4.75}$ & 55470.17$_{-4.85}^{+4.97}$  & 55214.75$_{-182.87}^{+146.26}$  \\[0.5ex]
$T_C$ & [d] & 55465.26$_{-4.08}^{+4.47}$ & 55469.18$_{-4.41}^{+4.30}$  & 55275.54$_{-121.34}^{+109.85}$  \\[0.5ex]
\hline\\[-1.5ex]
$Ar$ & [AU] & 0.324$_{-0.006}^{+0.006}$ & 0.323$_{-0.005}^{+0.006}$  & 1.036$_{-0.036}^{+0.051}$  \\[0.5ex]
M.$\sin{i}$ & [M$_{\rm Earth}$] & 30.56$_{-2.12}^{+2.10}$ & 31.29$_{-2.15}^{+2.20}$  & 27.26$_{-3.72}^{+3.84}$  \\[0.5ex]
\hline \\[-1.5ex]
$\gamma_{SOPHIE}$ & [m\,s$^{-1}$] & \multicolumn{1}{c}{-18737.35$_{-5.74}^{+5.38}$} & \multicolumn{2}{c}{-18715.26$_{-5.63}^{+5.58}$}\\[0.5ex]
$lin$ & [m\,s$^{-1}$\,yr$^{-1}$] & 7.71$_{-2.3}^{+2.42}$ & \multicolumn{2}{c}{-3.13$_{-2.34}^{+2.65}$}\\ [0.5ex]
$quad$& [m\,s$^{-1}$\,yr$^{-2}$] & -0.86$_{-0.25}^{+0.23}$ & \multicolumn{2}{c}{0.30$_{-0.26}^{+0.27}$} \\ [0.5ex]
$\sigma_{JIT}$ & [m\,s$^{-1}$] & \multicolumn{1}{c}{4.86$_{-0.62}^{+0.56}$} & \multicolumn{2}{c}{3.71$_{-0.46}^{+0.49}$}\\[0.5ex]
$\sigma_{(O-C)}$ & [m\,s$^{-1}$] & \multicolumn{1}{c}{4.55} & \multicolumn{2}{c}{3.90}\\[0.5ex]
$\log{(\rm Post})$ &   & \multicolumn{1}{c}{-466.65$_{-2.56}^{+1.82}$} & \multicolumn{2}{c}{-440.08$_{-3.07}^{+2.28}$}\\[0.5ex]
BIC &   & \multicolumn{1}{c}{105.93} & \multicolumn{2}{c}{118.52}\\[0.5ex]
\hline
\end{tabular}
\tablefoot{\tablefoottext{1}{Parameters obtained from the 1-keplerian model for Gl617A b.} \tablefoottext{2}{Parameters obtained from the 2-keplerian model for Gl617A b}.\\
For each parameter, the median of the posterior is reported, with error bars computed from the MCMC chains using a 68.3\% confidence interval. $\sigma_{O-C}$ corresponds to the weighted standard deviation of the residuals around this best solution. $\log{(\rm Post})$ is the posterior likelihood. All the parameters probed by the MCMC can be found in Appendix \ref{a:MCMC}, Tables \ref{Gl617A_tab-mcmc-Probed_params_k1d2} (1-keplerian model) and \ref{Gl617A_tab-mcmc-Probed_params_k2d2} (2-keplerian model).}
\end{table*}

\begin{figure}
\centering
\includegraphics[width=\hsize]{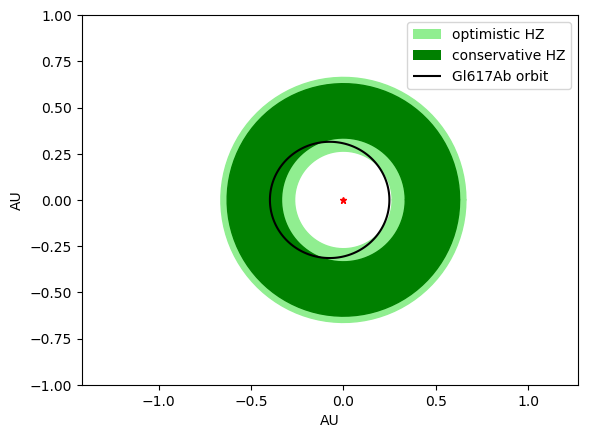}
\caption{Calculated orbit for Gl617A b with respect to the optimistic and conservative habitable zones, as defined by \cite{Kopparapu13}.}
\label{Gl617A_HZ}
\end{figure}

As for Gl96, we employed the habitable zone calculator to estimate the location of the HZ for Gl617A. The orbits calculated by the MCMC analysis place Gl617A b at 0.32AU, closer to the star than the conservative inner limit, but within the optimistic HZ (Figure \ref{Gl617A_HZ}). This differs from the results of \cite{Reiners}. As Gl617A b is moderately eccentric, we also calculate the mean incident flux (as done for Gl96 b) in order to better quantify its habitability. The mean incident flux is $<F>/F_ \oplus = 1.053$, placing it between the Recent Venus and Runaway Greenhouse limits as defined by \cite{Kopparapu13b}.

\subsubsection{Combination with CARMENES and KECK data}

\begin{figure}
\centering
    \includegraphics[width=\hsize]{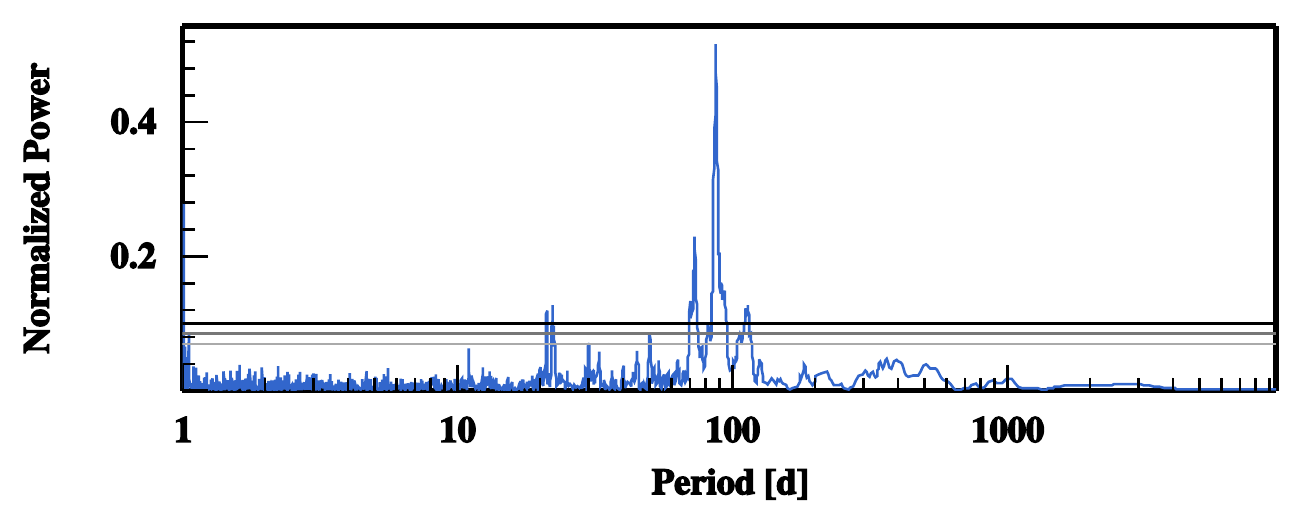}
\includegraphics[width=\hsize]{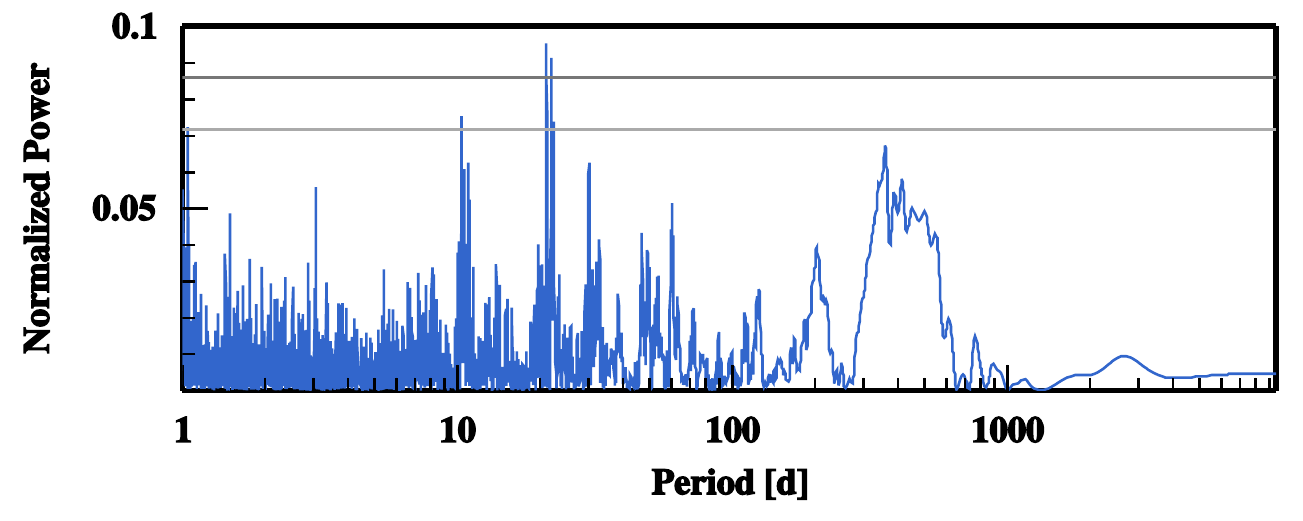}
\caption{Periodograms of: (top) the combined SOPHIE, CARMENES, and KECK radial velocities of Gl617A; (bottom) the residuals of a keplerian fit with P=86d to the data. The horizontal lines correspond to 50\%, 10\%, and (top periodogram only) 1\% FAP respectively.}
\label{Gl617A_alldat_per}
\end{figure}

We combined our observations of Gl617A with the CARMENES data presented by \cite{Reiners} and the KECK data of \cite{Butler}. Figure \ref{Gl617A_alldat_per} shows the resulting periodogram, with a strong signal at 86d, to which we fit a keplerian model using DACE.; and the periodogram of the residuals of this fit. The only strong signal is at 21d, and presumably corresponds to the stellar activity as discussed previously. The 500d signal that is present in the SOPHIE data is not in evidence in the combined observations, suggesting it may be spurious. 

\begin{figure}
\centering
\includegraphics[width=\hsize]{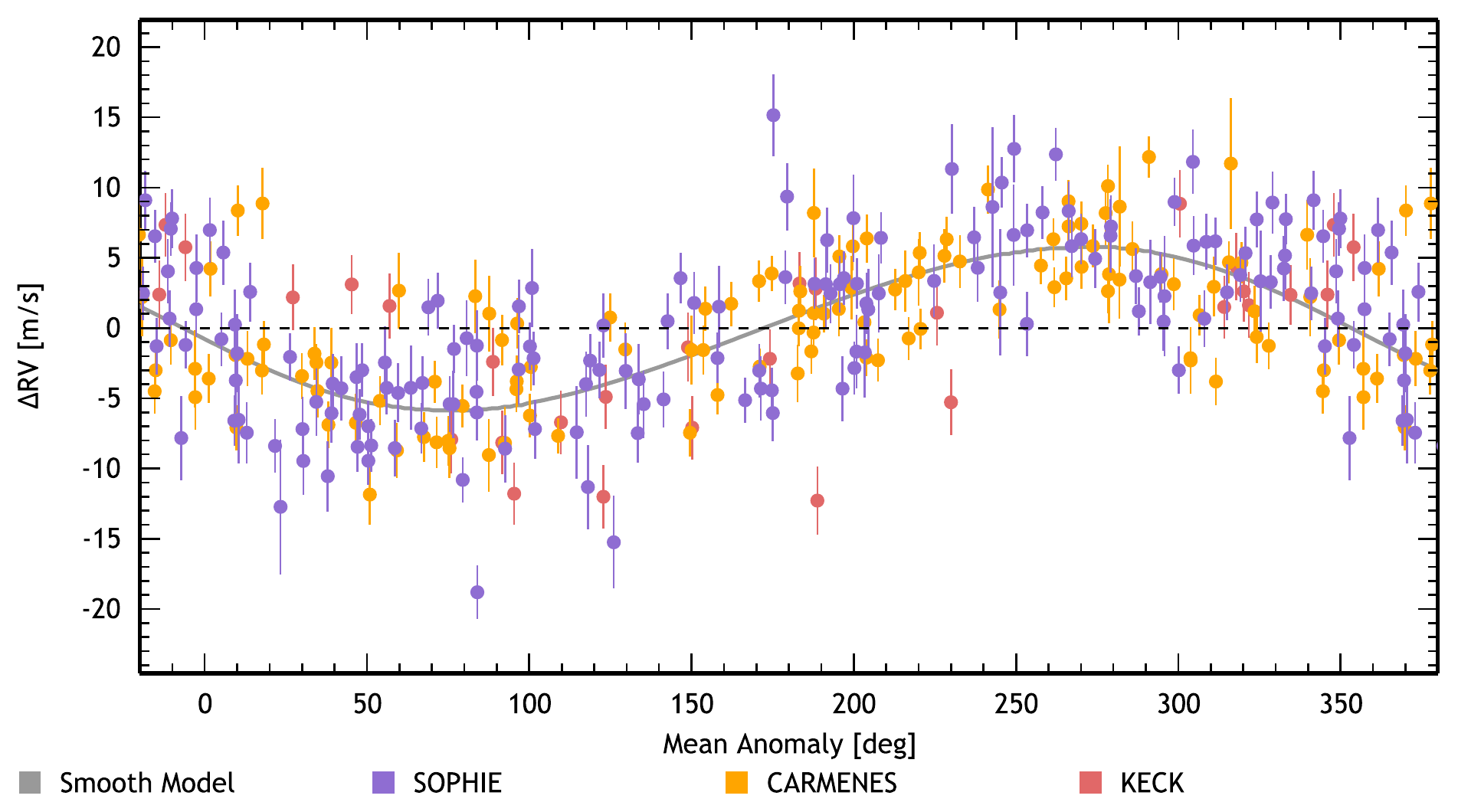}
\caption{Phase-folded radial velocities of Gl617A for a P=86d planet, using the combined SOPHIE, CARMENES, and KECK data.}
\label{Gl617A_alldat_phase}
\end{figure}

The posterior distribution of the model parameters was sampled using an MCMC algorithm implemented in DACE. A single-keplerian model with a quadratic drift was used. The resulting best-fit solution is summarised in Table \ref{Gl617A_alldat_tab-mcmc-Summary_params}, and the phase-folded combined data is shown in Figure \ref{Gl617A_alldat_phase}.

We compared the parameters resulting from the analysis of the combined data with those we obtained from SOPHIE data alone, and with those presented by \cite{Reiners}. In all three cases, the orbital periods are compatible at the 1$\sigma$ level (though slightly larger for the SOPHIE and combined data). M.$\sin{i}$ is also compatible at the 1$\sigma$ level between SOPHIE data alone and the combined data, but \cite{Reiners} present a slightly smaller value of M.$\sin{i}$ = 24.7$_{-1.8}^{+2.4}$ M$_{\rm Earth}$, only compatible with the others at the at the 2$\sigma$ level. The eccentricity is also different; the one derived from the SOPHIE data alone is larger, while those of the combined data and of \cite{Reiners} are compatible at 1$\sigma$. The amplitude of the signal, correspondingly, differs, being largest for the SOPHIE data alone and smallest in \cite{Reiners}'s analysis; this is coherent with the fact that for fixed $e$, $K$ grows with M.$\sin{i}$, and for fixed M.$\sin{i}$ $K$ grows with $e$ (see e.g. \citealt{Seager}).

Recently \cite{Feng} published a analysis of this system in the Research Note of the AAS, also employing the CARMENES, HIRES, and SOPHIE data. Their use of our SOPHIE data, however, is biased by the fact that they do not take into account the CTI effect and zero-point drift described in Section \ref{s:data} that we correct for in this work.

\begin{table}
\caption{Best-fitted solution for the planetary system orbiting Gl617A, from SOPHIE, CARMENES, and KECK combined data.}  \label{Gl617A_alldat_tab-mcmc-Summary_params}
\begin{tabular}{lcc}
\hline
\hline
Param. & Units & Gl617A b \\
\hline\\[-1.5ex]
$P$ & [d] & 86.78$_{-0.15}^{+0.16}$  \\[0.5ex]
$K$ & [m\,s$^{-1}$] & 5.83$_{-0.24}^{+0.22}$  \\[0.5ex]
$e$ &   & 0.07$_{-0.04}^{+0.04}$  \\[0.5ex]
$\omega$ & [deg] & 97.22$_{-41.83}^{+30.39}$  \\[0.5ex]
$T_P$ & [d] & 55468.26$_{-10.67}^{+8.52}$  \\[0.5ex]
$T_C$ & [d] & 55466.91$_{-4.11}^{+4.12}$  \\[0.5ex]
\hline \\[-1.5ex]
$Ar$ & [AU] & 0.323$_{-0.006}^{+0.005}$  \\[0.5ex]
M.$\sin{i}$ & [M$_{\rm Earth}$] & 28.55$_{-1.45}^{+1.49}$  \\[0.5ex]
\hline \\[-1.5ex]
$\gamma_{CARMENES}$ & [m\,s$^{-1}$] & \multicolumn{1}{c}{0.40$_{-1.60}^{+1.64}$}\\[0.5ex]
$\gamma_{KECK-PUB_1}$ & [m\,s$^{-1}$] & \multicolumn{1}{c}{-1.76$_{-2.68}^{+2.63}$}\\[0.5ex]
$\gamma_{SOPHIE}$ & [m\,s$^{-1}$] & \multicolumn{1}{c}{-18721.59$_{-1.50}^{+1.54}$}\\[0.5ex]
$\sigma_{JIT}$ & [m\,s$^{-1}$] & \multicolumn{1}{c}{15.89$_{-1.82}^{+4.14}$}\\[0.5ex]
$\sigma_{(O-C)}$ & [m\,s$^{-1}$] & \multicolumn{1}{c}{3.87}\\[0.5ex]
$\log{(\rm Post})$ &   & \multicolumn{1}{c}{-939.95$_{-2.76}^{+2.11}$}\\[0.5ex]
BIC &   & \multicolumn{1}{c}{1668.19}\\[0.5ex]
\hline
\end{tabular}
\tablefoot{Same notes as Table \ref{Gl617A_tab-mcmc-Summary_params_test}. All the parameters probed by the MCMC can be found in Appendix \ref{a:MCMC}, Table \ref{Gl617A_alldat_tab-mcmc-Probed_params}}
\end{table}

\section{Photometry}
\label{s:phot}

Stellar activity can be reflected not only in the radial velocity but in the photometric observations of a star, where we may also hope to find signals linked to the rotation period. In order to analyse whether this is the case for our targets, we obtained the Hipparcos photometry for both from \cite{Hipparcos}. 

\begin{figure}
\centering
 \includegraphics[width=\hsize]{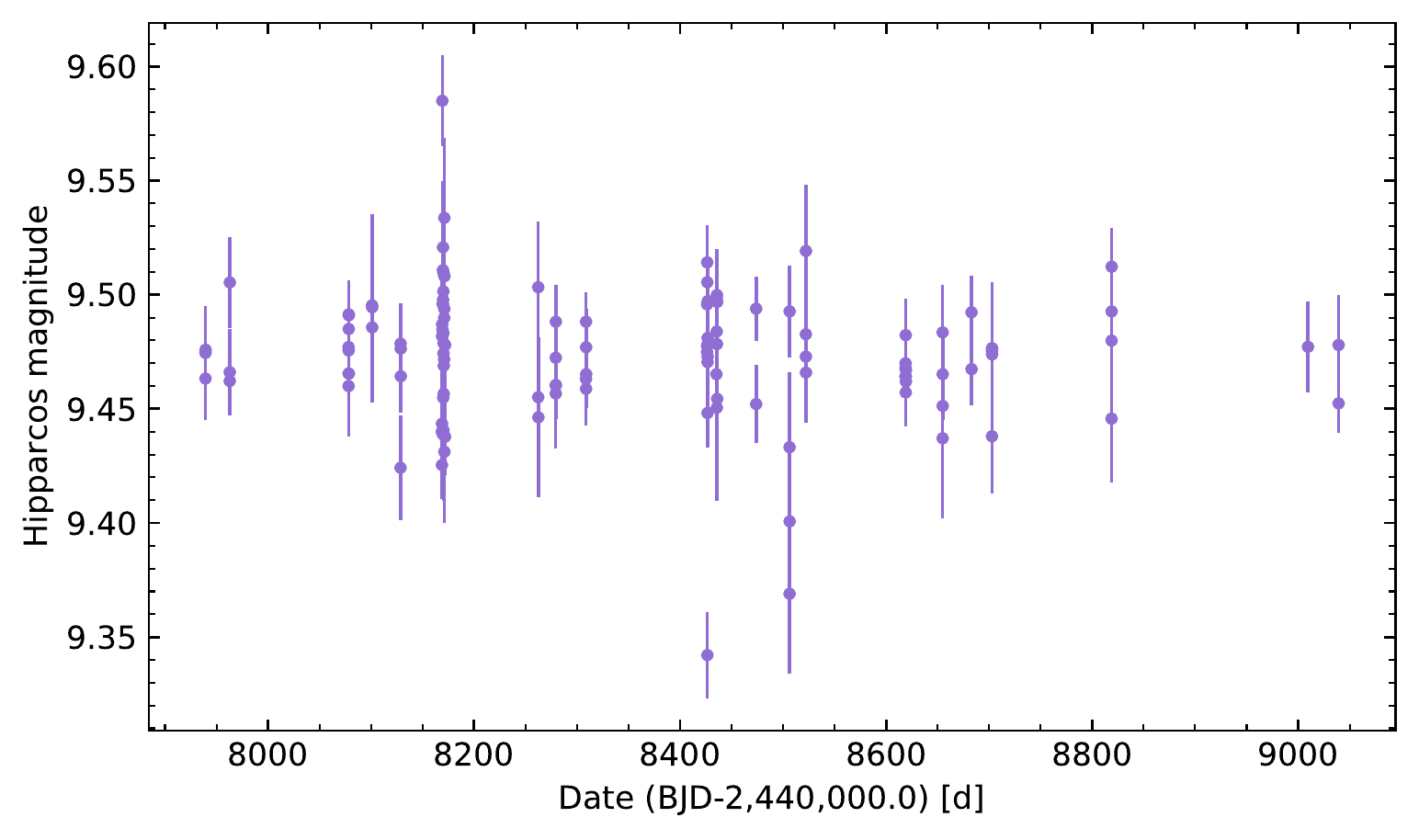}
 \includegraphics[width=\hsize]{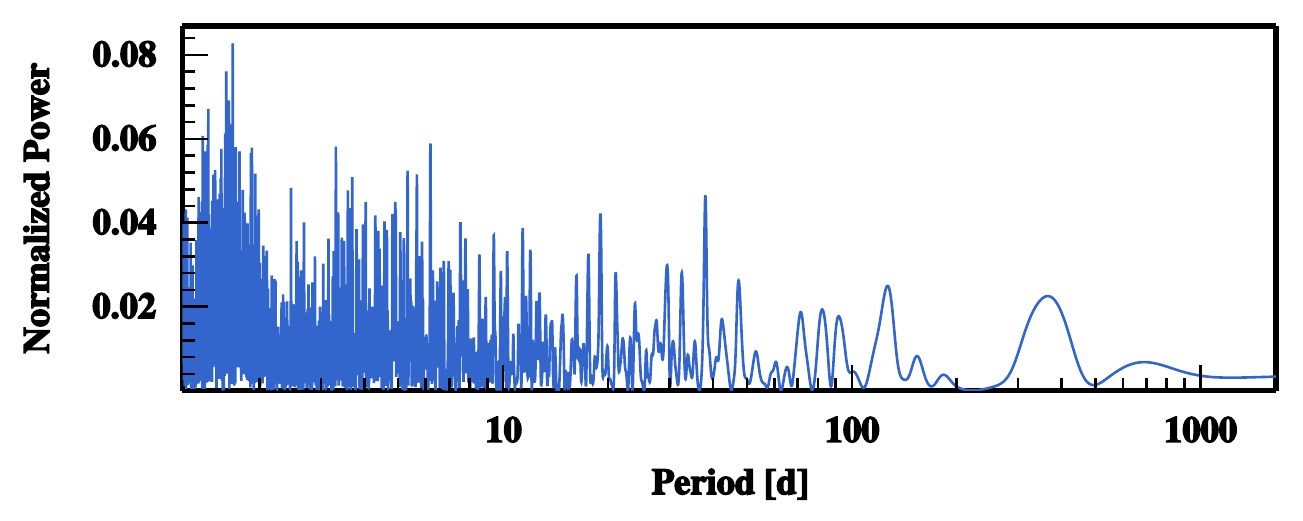}
\caption{Hipparcos photometry for Gl96 (top) and its corresponding periodogram (bottom).}
\label{Gl96_phot}
\end{figure}

For Gl96, we retrieved 121 measurements over a time span of three years. Fig. \ref{Gl96_phot} shows the photometric data points and the corresponding periodogram. No signals below 50\% FAP can be found in the periodogram. There is a small peak at 29d, the period found in the RV, H$\rm \alpha$, and $\rm log(R'_{HK})$ periodograms that is presumed related to activity, but it is not particularly relevant in the periodogram of photometric measurements which is dominated by a forest of peaks at around 2d. This set of peaks is probably linked to the sampling - as can be seen in the upper panel of Fig. \ref{Gl96_phot}, this star appears to have been observed in groups of several measurements within two days, with the groups set weeks or months apart.  

\begin{figure}
\centering
 \includegraphics[width=\hsize]{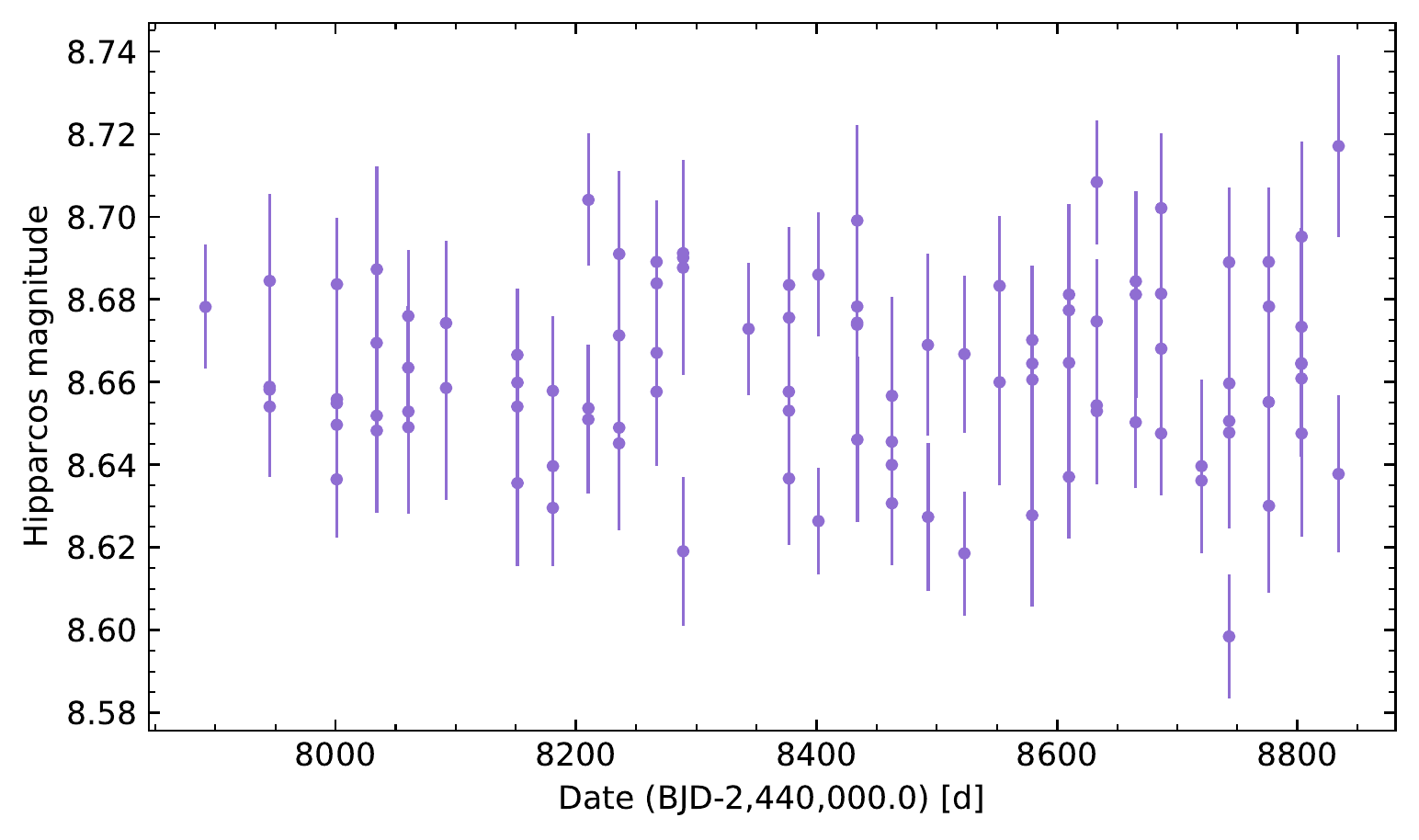}
 \includegraphics[width=\hsize]{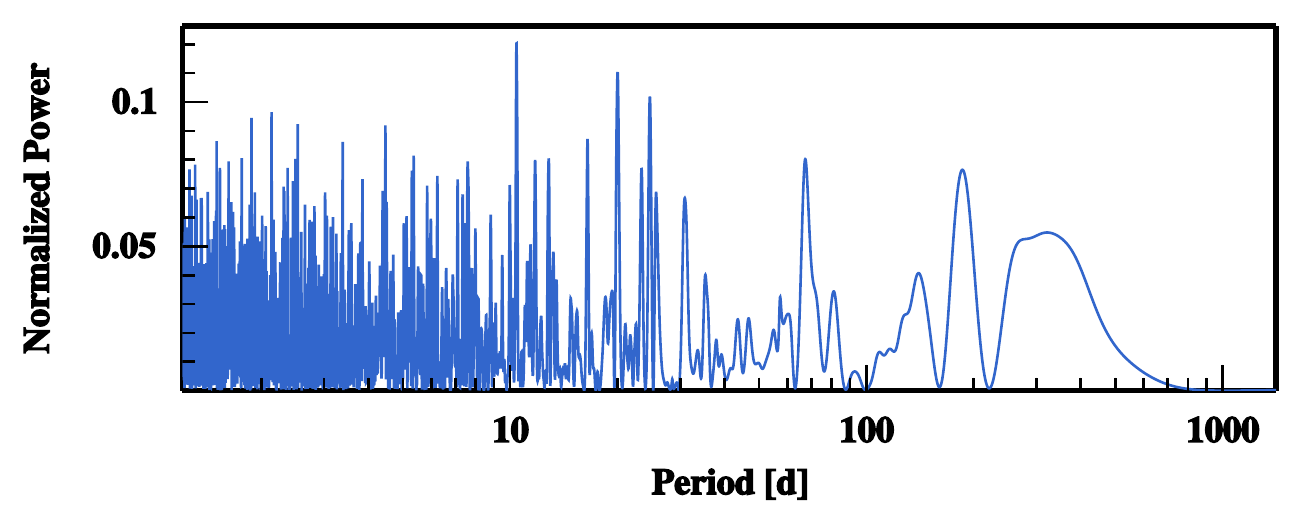}
\caption{Hipparcos photometry for Gl617A (top) and its corresponding periodogram (bottom).}
\label{Gl617A_phot}
\end{figure}

For Gl617A, we retrieved 103 measurements over a time span of two and a half years. The photometric data points and the corresponding periodogram are shown in Fig. \ref{Gl617A_phot}. No signals below 50\% FAP can be found in the periodogram. Nevertheless, it is worth remarking that the two highest peaks are at 10.4d and 20d; these values are very close to the 21d period seen in the RV and H$\rm \alpha$ periodograms and half this period.  

For both stars, we note that the HIPPARCOS photometry has a relatively high RMS (2.62\% for Gl96, 1.97\% for Gl617A). Activity signals at the level attained by our RV measurements would fall well below this, so any photometric activity tracers may be absorbed in the uncertainty of the data.

%--------------------------------------------------------------------

\section{Discussion and Conclusions}
\label{s:disc}

We have presented the detection of a new Neptune-like exoplanet orbiting the M dwarf Gl96 and the independent detection of a second Neptune-like exoplanet orbiting the M dwarf Gl617A for which we refine the planetary parameters, using the SOPHIE+ spectrograph on a 1.93m telescope. The planets have minimum masses of 29 and 31 Earth masses and orbital periods of 74 and 87 days respectively, and are located close to the inner limit of the HZ. For Gl96 we find no evidence for further planetary companions. Gl617A shows an intriguing signal of moderate significance at $\sim$500d in the periodogram that is best fit by a keplerian model. For Gl617A, we also analysed the combination of our data with that from CARMENES \citep{Reiners} and KECK \citep{Butler}. The 87d signal is clear in the combined data, though the resulting planetary parameters differ slightly from those obtained by SOPHIE data alone. The 500d signal, however, is no longer significant, which would suggest it is spurious. We may also suspect the influence of a magnetic cycle here; complementary observations in polarimetry with SPIRou should help to resolve the question.

\begin{figure}
\centering
\includegraphics[width=\hsize]{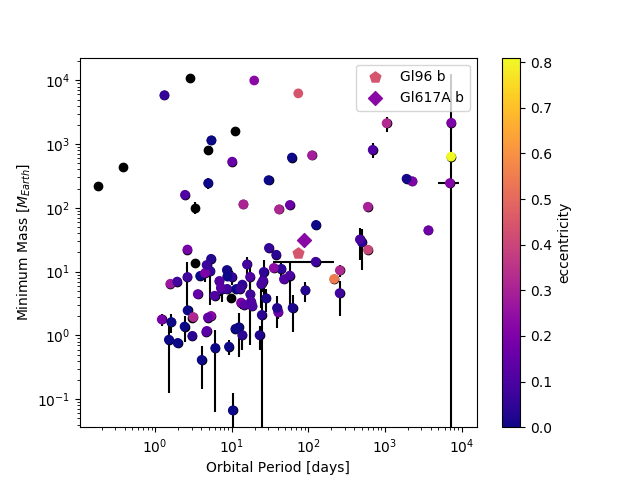}
\caption{Orbital period versus minimum mass for planets around M-dwarfs detected by radial velocities or transits. Each point represents a planet, with Gl96 b indicated by a pentagon and Gl617A b by a diamond. The colours indicate the orbital eccentricity.}
\label{exop_sample}
\end{figure}

As mentioned in Section \ref{s:int}, some 146 exoplanets around M-dwarf stars are presently known, of which 75 were detected by radial velocities and 33 by transits. The two planets presented here fall in the mid-to-long period and mid-to-high mass ranges of this sample, as shown in Figure \ref{exop_sample} (median period $= 13.5$ d, median mass $= 14.3 \ \rm M_{ Earth}$). Gl96 b is among the most eccentric known planets around M-dwarf stars (mean e $= 0.12$), surpassed only by Wolf 1061 d \citep{Wright} and GJ 317 c \citep{Johnson}, with eccentricities of 0.55 and 0.81 respectively.

Both host stars are metal-rich, as noted in Table \ref{starprop}. This is consistent with a trend found for M-dwarfs hosting planets to be preferentially metal-rich (e.g. \citealt{Courcol16}, \citealt{Hobson}). Additionally, the masses determined for the planets are compatible with the upper mass boundary determined for Neptune-like planets around M-dwarfs by \cite{Courcol16}, which corresponds to around 35 Earth masses for the metallicities of these stars.

Low-mass planets orbiting M-dwarfs are mainly found in multi-planet systems. Follow-up observations with SPIRou in spectropolarimetry will permit us to disentangle the stellar activity and planetary signals, refine the mass and identify possible additional rocky planets.

%--------------------------------------------------------------------

\begin{acknowledgements}
We warmly thank the OHP staff for their support on the observations.
XD, XB, IB and TF received funding from the French Programme
National de Physique Stellaire (PNPS) and the Programme National
de Planétologie (PNP) of CNRS (INSU).
NCS and OD were supported by Funda\c{c}\~ao para a Ci\^encia e a Tecnologia (FCT, Portugal) through national funds and by FEDER through COMPETE2020 in the context of the projects and grants reference UID/FIS/04434/2013 \& POCI-01-0145-FEDER-007672, PTDC/FIS-AST/1526/2014 \& POCI-01-0145-FEDER-016886, and IF/00169/2012/CP0150/CT0002.
This publication makes use of The Data \& Analysis Center for Exoplanets (DACE), which is a facility based at the University of Geneva (CH) dedicated to extrasolar planets data visualisation, exchange and analysis. DACE is a platform of the Swiss National Centre of Competence in Research (NCCR) PlanetS, federating the Swiss expertise in Exoplanet research. The DACE platform is available at https://dace.unige.ch.
We thank the anonymous referees for their comments which helped to improve the manuscript.
\end{acknowledgements}

%\begin{thebibliography}{}
\bibliographystyle{aa} % style aa.bst
\bibliography{output.bbl} % your references Yourfile.bib
%\end{thebibliography}

%-------------------------------------------------------------------

\begin{appendix}

\section{Radial Velocities}
\label{A:radvel}
 
 In this Appendix, we present for each star the full set of radial velocities derived by template-matching, with the short-term instrumental drift and long-term zero-point corrections applied.
 
 \clearpage
\longtab[1]{ 
\begin{longtable}{lll}
\caption{\label{Gl96_RV_t} Radial velocities for Gl96 derived by template-matching, with the short-term instrumental drift and long-term zero-point corrections applied.}\\
\hline\hline
BJD $[-2400000 d]$ & RV $[km/s]$ & sigma RV $[km/s]$ \\
\hline
\endfirsthead
\caption{continued.}\\
\hline\hline
BJD $[-2400000 d]$ & RV $[km/s]$ & sigma RV $[km/s]$ \\
\hline
\endhead
\hline
\endfoot
55813.6543 & -37.8703 & 0.0023 \\
55916.385  & -37.8778 & 0.003  \\
56177.6458 & -37.8799 & 0.002  \\
57260.6283 & -37.8687 & 0.0018 \\
57261.6026 & -37.8647 & 0.0017 \\
57262.591  & -37.8692 & 0.0017 \\
57263.5689 & -37.8714 & 0.0018 \\
57264.6339 & -37.8793 & 0.0019 \\
57269.613  & -37.8702 & 0.0021 \\
57272.5499 & -37.8704 & 0.003  \\
57285.486  & -37.8727 & 0.0023 \\
57287.5502 & -37.8739 & 0.0018 \\
57289.513  & -37.8721 & 0.0025 \\
57291.5016 & -37.8723 & 0.0023 \\
57292.6058 & -37.88   & 0.0017 \\
57318.4756 & -37.8718 & 0.002  \\
57319.5631 & -37.8764 & 0.0021 \\
57324.5429 & -37.8759 & 0.0023 \\
57325.525  & -37.8747 & 0.0017 \\
57326.4449 & -37.8705 & 0.0017 \\
57327.4813 & -37.8652 & 0.0018 \\
57341.5201 & -37.8722 & 0.0023 \\
57342.4705 & -37.8749 & 0.0018 \\
57349.3667 & -37.875  & 0.002  \\
57352.4482 & -37.8775 & 0.0026 \\
57353.411  & -37.8796 & 0.0021 \\
57373.4177 & -37.8781 & 0.0018 \\
57375.413  & -37.8766 & 0.002  \\
57631.6103 & -37.867  & 0.002  \\
57632.6444 & -37.8706 & 0.0023 \\
57634.637  & -37.8721 & 0.0021 \\
57635.6266 & -37.8683 & 0.0018 \\
57639.5888 & -37.8729 & 0.0021 \\
57640.5853 & -37.8747 & 0.002  \\
57642.596  & -37.8722 & 0.0024 \\
57652.5243 & -37.8814 & 0.0024 \\
57656.5486 & -37.8768 & 0.0025 \\
57657.5581 & -37.8723 & 0.0025 \\
57664.4734 & -37.8716 & 0.0024 \\
57665.4864 & -37.875  & 0.0022 \\
57666.5107 & -37.8758 & 0.0017 \\
57667.5688 & -37.8812 & 0.0027 \\
57669.5344 & -37.8772 & 0.0021 \\
57670.4846 & -37.8803 & 0.0024 \\
57671.4838 & -37.8748 & 0.0025 \\
57677.513  & -37.8789 & 0.0016 \\
57680.5849 & -37.8735 & 0.0022 \\
57700.6179 & -37.8657 & 0.0022 \\
57701.4932 & -37.8712 & 0.0024 \\
57709.5002 & -37.874  & 0.0017 \\
57712.2714 & -37.8743 & 0.0026 \\
57729.3458 & -37.8747 & 0.0017 \\
57730.4299 & -37.8755 & 0.0017 \\
57731.4116 & -37.8782 & 0.0017 \\
57732.4296 & -37.8821 & 0.0018 \\
57733.3579 & -37.8835 & 0.0017 \\
57734.3638 & -37.8815 & 0.0024 \\
57740.4115 & -37.8775 & 0.0019 \\
57741.3933 & -37.8712 & 0.0019 \\
57758.299  & -37.875  & 0.0025 \\
57761.2904 & -37.8822 & 0.0023 \\
57762.3098 & -37.8812 & 0.0017 \\
57765.2701 & -37.8794 & 0.0019 \\
57770.3366 & -37.8726 & 0.0027 \\
57772.3558 & -37.8708 & 0.0017 \\
57801.3467 & -37.8733 & 0.0031 \\
57815.3111 & -37.8793 & 0.0019 \\
57822.3095 & -37.885  & 0.0029 \\
57823.3328 & -37.8796 & 0.0021 \\
57825.2954 & -37.8803 & 0.0024 \\
57828.2853 & -37.877  & 0.0022 \\
57832.2892 & -37.8687 & 0.0019  \\
\end{longtable}
}

\clearpage
\longtab[2]{ 
\begin{longtable}{lll}
\caption{\label{Gl617A_RV_t} Radial velocities for Gl617A derived by template-matching, with the short-term instrumental drift and long-term zero-point corrections applied.}\\ 
\hline\hline
BJD $[-2400000 d]$ & RV $[km/s]$ & sigma RV $[km/s]$ \\
\hline
\endfirsthead
\caption{continued.}\\
\hline\hline
BJD $[-2400000 d]$ & RV $[km/s]$ & sigma RV $[km/s]$ \\
\hline
\endhead
\hline
\endfoot
55827.3208           & -18.7247      & 0.0019              \\
55982.7022           & -18.714       & 0.0018              \\
56328.6907           & -18.7186      & 0.0019              \\
56370.663            & -18.7214      & 0.002               \\
56372.632            & -18.7201      & 0.0022              \\
56377.596            & -18.7262      & 0.0023              \\
56405.4505           & -18.7182      & 0.002               \\
56419.4838           & -18.7232      & 0.002               \\
56425.4627           & -18.7237      & 0.0028              \\
56469.5016           & -18.4846      & 0.0465              \\
56471.4651           & -18.7236      & 0.0026              \\
56487.4569           & -18.7161      & 0.0022              \\
56488.4484           & -18.7198      & 0.0056              \\
56490.3756           & -18.7147      & 0.0022              \\
56496.4612           & -18.7101      & 0.0023              \\
56497.4503           & -18.7158      & 0.002               \\
56500.3679           & -18.7166      & 0.0019              \\
56501.48             & -18.7186      & 0.0036              \\
56502.3675           & -18.713       & 0.0022              \\
56519.3387           & -18.7259      & 0.0021              \\
56521.3385           & -18.7281      & 0.0018              \\
56523.3515           & -18.7262      & 0.0019              \\
56556.3177           & -18.7195      & 0.0021              \\
56557.3098           & -18.7184      & 0.0019              \\
56558.2799           & -18.7188      & 0.0018              \\
56583.2842           & -18.7161      & 0.0021              \\
56613.2241           & -18.7205      & 0.002               \\
56624.2446           & -18.7294      & 0.0033              \\
56625.2369           & -18.7243      & 0.0019              \\
56626.2231           & -18.7218      & 0.0023              \\
56629.2163           & -18.7274      & 0.0024              \\
56694.6636           & -18.7255      & 0.0024              \\
56705.6963           & -18.7306      & 0.0024              \\
56711.7062           & -18.726       & 0.0023              \\
56712.6984           & -18.725       & 0.002               \\
56730.5614           & -18.7189      & 0.002               \\
56732.6738           & -18.7207      & 0.0029              \\
56733.6147           & -18.7156      & 0.0018              \\
56761.5278           & -18.7143      & 0.002               \\
56777.4867           & -18.7315      & 0.0024              \\
56779.5828           & -18.7281      & 0.0021              \\
56782.5313           & -18.7304      & 0.0019              \\
56793.4986           & -18.7205      & 0.0019              \\
56794.5888           & -18.7242      & 0.0019              \\
56800.5492           & -18.7373      & 0.0033              \\
56815.4608           & -18.7189      & 0.0018              \\
56818.4207           & -18.7249      & 0.0019              \\
56835.3756           & -18.7293      & 0.0058              \\
56841.3896           & -18.7216      & 0.0021              \\
56844.4113           & -18.7214      & 0.002               \\
56850.4049           & -18.7169      & 0.0024              \\
56854.42             & -18.715       & 0.0019              \\
56859.4538           & -18.7286      & 0.0031              \\
56862.5723           & -18.7348      & 0.0048              \\
56876.3988           & -18.7228      & 0.0027              \\
56885.4011           & -18.7334      & 0.003               \\
56904.3149           & -18.7264      & 0.0023              \\
56914.3578           & -18.7178      & 0.0024              \\
56926.3241           & -18.7209      & 0.0017              \\
56929.2964           & -18.7251      & 0.0017              \\
56939.3032           & -18.713       & 0.0021              \\
56941.2947           & -18.7143      & 0.0021              \\
56942.2937           & -18.7233      & 0.0017              \\
56962.2331           & -18.7236      & 0.0017              \\
56968.2461           & -18.7293      & 0.0021              \\
56999.2152           & -18.7108      & 0.0032              \\
57002.2308           & -18.7135      & 0.0057              \\
57072.6911           & -18.7282      & 0.002               \\
57079.64             & -18.7204      & 0.0022              \\
57084.6695           & -18.7188      & 0.0026              \\
57087.6397           & -18.7157      & 0.0022              \\
57089.6958           & -18.7118      & 0.0018              \\
57090.615            & -18.7094      & 0.0024              \\
57091.5817           & -18.7152      & 0.0019              \\
57115.5282           & -18.73        & 0.003               \\
57119.5769           & -18.7259      & 0.0023              \\
57438.6677           & -18.722       & 0.0023              \\
57465.6133           & -18.7231      & 0.0019              \\
57466.5738           & -18.7289      & 0.0018              \\
57469.5975           & -18.7307      & 0.0019              \\
57471.637            & -18.7295      & 0.0023              \\
57474.5271           & -18.7266      & 0.0023              \\
57476.5225           & -18.7293      & 0.0018              \\
57496.5362           & -18.7298      & 0.0021              \\
57498.4581           & -18.7274      & 0.002               \\
57502.5692           & -18.7208      & 0.0029              \\
57507.5044           & -18.7187      & 0.0019              \\
57510.5087           & -18.7192      & 0.0031              \\
57512.5605           & -18.7145      & 0.0031              \\
57521.4872           & -18.4686      & 0.0648              \\
57523.4672           & -18.7198      & 0.0045              \\
57524.4615           & -18.7157      & 0.0036              \\
57526.6088           & -18.7141      & 0.0019              \\
57528.5508           & -18.714       & 0.0021              \\
57529.4749           & -18.716       & 0.0018              \\
57530.5157           & -18.7174      & 0.0021              \\
57544.5369           & -18.7181      & 0.0023              \\
57547.4552           & -18.7158      & 0.0019              \\
57548.4152           & -18.7183      & 0.0023              \\
57549.4331           & -18.4442      & 0.0485              \\
57550.5206           & -18.721       & 0.002               \\
57553.3989           & -18.7221      & 0.0025              \\
57557.4901           & -18.7244      & 0.0017              \\
57559.4385           & -18.7276      & 0.0024              \\
57562.4873           & -18.4644      & 0.0478              \\
57562.5082           & -18.7308      & 0.0018              \\
57564.5346           & -18.7248      & 0.0026              \\
57566.4621           & -18.7266      & 0.003               \\
57568.4546           & -18.7204      & 0.002               \\
57571.3651           & -18.7236      & 0.0025              \\
57574.4738           & -18.7253      & 0.0025              \\
57575.4626           & -18.7195      & 0.0028              \\
57582.4238           & -18.7254      & 0.0027              \\
57583.3905           & -18.726       & 0.0025              \\
57586.4968           & -18.7188      & 0.0018              \\
57593.393            & -18.7072      & 0.0029              \\
57594.408            & -18.713       & 0.0024              \\
57597.3821           & -18.7161      & 0.0023              \\
57614.3752           & -18.71        & 0.0019              \\
57618.4374           & -18.7158      & 0.0026              \\
57621.3838           & -18.7191      & 0.003               \\
57622.4412           & -18.7201      & 0.0043              \\
57633.3421           & -18.7199      & 0.0019              \\
57635.32             & -18.7217      & 0.002               \\
57637.3095           & -18.7181      & 0.0024              \\
57638.307            & -18.7154      & 0.0023              \\
57639.2986           & -18.4182      & 0.0437              \\
57639.3109           & -18.717       & 0.0023              \\
57641.3044           & -18.7198      & 0.0021              \\
57652.2945           & -18.727       & 0.0021              \\
57656.3939           & -18.7278      & 0.0024              \\
57795.702            & -18.7188      & 0.0018              \\
57800.6382           & -18.7199      & 0.002               \\
57814.6062           & -18.7299      & 0.0022              \\
57820.6203           & -18.733       & 0.0025              \\
57823.6354           & -18.7319      & 0.0017              \\
57825.6004           & -18.731       & 0.002               \\
57827.5652           & -18.7296      & 0.0018              \\
57831.6704           & -18.727       & 0.0023              \\
57849.5691           & -18.7246      & 0.0018              \\
57851.6214           & -18.7276      & 0.0016              \\
57852.6711           & -18.7255      & 0.0019              \\
57853.6085           & -18.7269      & 0.0016              \\
57860.5126           & -18.7242      & 0.0032              \\
57861.5489           & -18.72        & 0.0028              \\
57883.518            & -18.7135      & 0.0017              \\
57886.5506           & -18.7163      & 0.0017              \\
57888.4005           & -18.7187      & 0.0019              \\
57914.4292           & -18.7264      & 0.0019              \\
57916.4513           & -18.7279      & 0.002               \\
57917.4221           & -18.7333      & 0.0016              \\
57918.4669           & -18.7285      & 0.002               \\
57918.5051           & -18.7413      & 0.0019              \\
57922.4145           & -18.7238      & 0.0017              \\ 
%\end{tabular}
\end{longtable}
}

\section{MCMC analysis - Full Probed Parameters}
\label{a:MCMC}

\subsection{Gl96}

We present the full set of parameters probed by the DACE MCMC analysis for Gl96, and the derived physical parameters.

\begin{table*}
\caption{Parameters probed by the MCMC used to fit the RV measurements of Gl96.}  \label{Gl96_tab-mcmc-Probed_params}
\tiny
\begin{center}
\begin{tabular}{lcclcccccccc}
\hline
\hline
Param. & Units & Max(Like) & Med & Mod &Std & CI(15.85) & CI(84.15) &CI(2.275) & CI(97.725) & Prior\\
\hline
\multicolumn{11}{c}{ \bf Likelihood}\\
\hline
$\log{(\rm Post})$&               &  -193.285494&  -196.771970&  -195.921884&     1.870757&  -199.232726&  -195.096198&  -202.542797&  -194.075043&               \\ 
$\log{(\rm Like)}$&               &  -191.922894&  -195.623787&  -195.175969&     1.973270&  -198.178506&  -193.846439&  -201.754278&  -192.754170&               \\ 
$\log{(\rm Prior)}$&               &    -1.362600&    -1.078578&    -1.136538&     0.427678&    -1.579763&    -0.619581&    -2.223648&    -0.226630&               \\ 
\hline 
M$_{\star}$    &[M$_{\odot}$]  &     0.621188&     0.600424&     0.596743&     0.026436&     0.570211&     0.630397&     0.540296&     0.660160&$\mathcal{U}$  \\ 
\hline 
$\sigma_{JIT}$ &[m\,s$^{-1}$]  &     3.00&     3.45&     3.21&     0.93&     2.54&     2.30&     1.78&     0.77&$\mathcal{U}$  \\ 
\hline 
$\gamma_{SOPHIE}$&[m\,s$^{-1}$]  &-37874.88&-37874.84&-37874.90&     0.28&-37875.15&-37874.53&-37875.50&-37874.22&$\mathcal{U}$  \\ 
\hline 
\hline 
$P$            &[d]            &    73.939773&    73.937730&    73.993314&     0.325983&    73.553342&    74.264366&    73.092750&    74.719936&$\mathcal{U}$  \\ 
$K$            &[m\,s$^{-1}$]  &     5.20&     4.69&     4.42&     0.60&     4.07&     5.41&     3.50&     6.26&$\mathcal{U}$  \\ 
$e$            &               &     0.495245&     0.439034&     0.430347&     0.088190&     0.332797&     0.529659&     0.200769&     0.609855&$\mathcal{U}$  \\ 
$\omega$       &[deg]          &   341.181282&   339.576769&   338.551057&    11.790040&   325.056172&   352.022293&   307.050528&   358.379033&$\mathcal{U}$  \\ 
$\lambda_{0}$  &[deg]          &    64.471037&    63.383026&    67.891413&    45.293155&     9.598258&   108.428683&   -55.974137&   170.848142&$\mathcal{U}$  \\ 
\hline 
\end{tabular}
\tablefoot{The maximum likelihood solution (Max(Like)), the median (Med), mode (Mod) and standard deviation (Std) of the posterior distribution for each parameter is shown, as well as the 68.3\% (CI(15.85),CI(84.15)) and 95.45\% (CI(2.275),CI(97.725)) confidence intervals. The prior for each parameter can be of type: $\mathcal{U}$: uniform, $\mathcal{N}$: normal, $\mathcal{SN}$:split normal, $\mathcal{TN}$: truncated normal.}
\end{center}
\end{table*}

\begin{table*}
\caption{Physical parameters derived from the MCMC chains used fit the RV measurements of Gl96.}  \label{Gl96_tab-mcmc-Physical_params}
\tiny
\begin{center}
\begin{tabular}{lcclcccccccc}
\hline
\hline
Param. & Units & Max(Like) & Med & Mod &Std & CI(15.85) & CI(84.15) &CI(2.275) & CI(97.725) & Prior\\
\hline
\multicolumn{10}{c}{ \bf Likelihood}\\
\hline
M$_{\star}$    &[M$_{\odot}$]  &     0.621188&     0.600424&     0.596743&     0.026436&     0.570211&     0.630397&     0.540296&     0.660160&$\mathcal{U}$  \\ 
\hline 
$P$            &[d]            &    73.939773&    73.937730&    73.993314&     0.325983&    73.553342&    74.264366&    73.092750&    74.719936&$\mathcal{U}$  \\ 
$K$            &[m\,s$^{-1}$]  &     5.20&     4.69&     4.42&     0.60&     4.07&     5.41&     3.50&     6.26&$\mathcal{U}$  \\ 
$e$            &               &     0.495245&     0.439034&     0.430347&     0.088190&     0.332797&     0.529659&     0.200769&     0.609855&$\mathcal{U}$  \\ 
$\omega$       &[deg]          &   341.181282&   339.576769&   338.551057&    11.790040&   325.056172&   352.022293&   307.050528&   358.379033&$\mathcal{U}$  \\ 
$T_P$          &[d]            & 55556.833035& 55556.392386& 55553.872826&     9.006008& 55547.408500& 55566.959102& 55534.517800& 55579.539962&               \\ 
$T_C$          &[d]            & 55567.178726& 55568.899883& 55564.573169&     9.663639& 55559.078770& 55580.463887& 55546.269103& 55593.191399&               \\ 
\hline 
$Ar$           &[AU]           &     0.294178&     0.290799&     0.290868&     0.004383&     0.285732&     0.295728&     0.280513&     0.300435&               \\ 
M.$\sin{i}$    &[M$_{\rm Jup}$]&     0.067975&     0.061855&     0.058986&     0.006479&     0.054626&     0.069477&     0.047715&     0.077017&               \\ 
M.$\sin{i}$    &[M$_{\rm Earth}$]&    21.602453&    19.657445&    18.745664&     2.058966&    17.359998&    22.079752&    15.163668&    24.475879&               \\ 
\hline 
\end{tabular}
\tablefoot{The maximum likelihood solution (Max(Like)), the median (Med), mode (Mod) and standard deviation (Std) for the posterior distribution of each parameter is shown, as well as the 68.3\% (CI(15.85),CI(84.15)) and 95.45\% (CI(2.275),CI(97.725)) confidence intervals. The prior for each parameter can be of type: $\mathcal{U}$: uniform, $\mathcal{N}$: normal, $\mathcal{SN}$:split normal, $\mathcal{TN}$: truncated normal.}
\end{center}
\end{table*}

\subsection{Gl617A}

We present the full set of parameters probed by the DACE MCMC analysis for Gl617A, and the derived physical parameters.

\begin{table*}
\caption{Parameters probed by the MCMC used to fit the RV measurements of Gl617A - 1-keplerian model plus quadratic drift.}   \label{Gl617A_tab-mcmc-Probed_params_k1d2}
\tiny
\begin{center}
\begin{tabular}{lcclcccccccc}
\hline
\hline
Param. & Units & Max(Like) & Med & Mod &Std & CI(15.85) & CI(84.15) &CI(2.275) & CI(97.725) & Prior\\
\hline
\multicolumn{11}{c}{ \bf Likelihood}\\
\hline
$\log{(\rm Post})$&               &  -462.619011&  -466.651637&  -466.520190&     1.964561&  -469.209353&  -464.831362&  -472.561763&  -463.627502&               \\ 
$\log{(\rm Like)}$&               &  -461.941145&  -466.049364&  -465.293525&     2.015104&  -468.693270&  -464.197613&  -472.107721&  -462.923059&               \\ 
$\log{(\rm Prior)}$&               &    -0.677866&    -0.555253&    -0.500711&     0.254783&    -0.874032&    -0.292873&    -1.220718&    -0.097352&               \\ 
\hline 
M$_{\star}$    &[M$_{\odot}$]  &     0.611414&     0.600358&     0.595467&     0.026063&     0.570942&     0.630052&     0.540715&     0.658600&$\mathcal{U}$  \\ 
\hline 
$\sigma_{JIT}$ &[m\,s$^{-1}$]  &     4.44&     4.86&     4.73&     0.52&     4.30&     4.24&     3.78&     3.55&$\mathcal{U}$  \\ 
\hline 
$\gamma_{SOPHIE}$&[m\,s$^{-1}$]  &-18739.60&-18737.35&-18736.94&     4.89&-18743.09&-18731.97&-18748.80&-18726.19&$\mathcal{U}$  \\ 
\hline 
$lin$          &[m\,s$^{-1}$\,yr$^{-1}$]&     8.53&     7.71&     7.09&     2.09&     5.41&    10.13&     3.01&    12.66&$\mathcal{U}$  \\ 
$quad$         &[m\,s$^{-1}$\,yr$^{-2}$]&    -0.93&    -0.86&    -0.84&     0.21&    -1.11&    -0.63&    -1.36&    -0.38&$\mathcal{U}$  \\ 
\hline 
$P$            &[d]            &    86.916721&    86.925044&    86.904924&     0.164238&    86.730726&    87.102317&    86.523585&    87.278783&$\mathcal{U}$  \\ 
$K$            &[m\,s$^{-1}$]  &     6.70&     6.56&     6.47&     0.35&     6.16&     6.97&     5.78&     7.37&$\mathcal{U}$  \\ 
$e$            &               &     0.349308&     0.316142&     0.319096&     0.073835&     0.229602&     0.396643&     0.132376&     0.468753&$\mathcal{U}$  \\ 
$\omega$       &[deg]          &   101.004334&   102.200583&   103.915448&     9.807382&    90.720509&   112.245164&    75.691327&   123.345529&$\mathcal{U}$  \\ 
$\lambda_{0}$  &[deg]          &   238.987027&   239.918335&   240.420912&    16.482165&   220.387042&   257.567720&   199.451375&   275.038557&$\mathcal{U}$  \\ \hline 
\end{tabular}
\tablefoot{The maximum likelihood solution (Max(Like)), the median (Med), mode (Mod) and standard deviation (Std) of the posterior distribution for each parameter is shown, as well as the 68.3\% (CI(15.85),CI(84.15)) and 95.45\% (CI(2.275),CI(97.725)) confidence intervals. The prior for each parameter can be of type: $\mathcal{U}$: uniform, $\mathcal{N}$: normal, $\mathcal{SN}$:split normal, $\mathcal{TN}$: truncated normal.}
\end{center}
\end{table*}

\begin{table*}
\caption{Physical parameters derived from the MCMC chains used fit the RV measurements of Gl617A - 1-keplerian model plus quadratic drift.}  \label{Gl617A_tab-mcmc-Physical_params_k1d2}
\tiny
\begin{center}
\begin{tabular}{lcclcccccccc}
\hline
\hline
Param. & Units & Max(Like) & Med & Mod &Std & CI(15.85) & CI(84.15) &CI(2.275) & CI(97.725) & Prior\\
\hline
\multicolumn{10}{c}{ \bf Likelihood}\\
\hline
M$_{\star}$    &[M$_{\odot}$]  &     0.611414&     0.600358&     0.595467&     0.026063&     0.570942&     0.630052&     0.540715&     0.658600&$\mathcal{U}$  \\ 
\hline 
$P$            &[d]            &    86.916721&    86.925044&    86.904924&     0.164238&    86.730726&    87.102317&    86.523585&    87.278783&$\mathcal{U}$  \\ 
$K$            &[m\,s$^{-1}$]  &     6.70&     6.56&     6.47&     0.35&     6.16&     6.97&     5.78&     7.37&$\mathcal{U}$  \\ 
$e$            &               &     0.349308&     0.316142&     0.319096&     0.073835&     0.229602&     0.396643&     0.132376&     0.468753&$\mathcal{U}$  \\ 
$\omega$       &[deg]          &   101.004334&   102.200583&   103.915448&     9.807382&    90.720509&   112.245164&    75.691327&   123.345529&$\mathcal{U}$  \\ 
$T_P$          &[d]            & 55466.686102& 55466.601086& 55465.672814&     4.040069& 55462.278510& 55471.350467& 55457.295504& 55476.587804&               \\ 
$T_C$          &[d]            & 55465.481743& 55465.258184& 55464.278744&     3.805920& 55461.181843& 55469.727111& 55456.842749& 55474.584025&               \\ 
\hline 
$Ar$           &[AU]           &     0.325940&     0.323965&     0.324425&     0.004713&     0.318459&     0.329242&     0.312843&     0.334318&               \\ 
M.$\sin{i}$    &[M$_{\rm Jup}$]&     0.098576&     0.096167&     0.094936&     0.005835&     0.089490&     0.102775&     0.083150&     0.110012&               \\ 
M.$\sin{i}$    &[M$_{\rm Earth}$]&    31.327394&    30.561882&    30.170671&     1.854279&    28.440031&    32.661895&    26.425117&    34.961725&               \\ 
\hline 
\end{tabular}
\tablefoot{The maximum likelihood solution (Max(Like)), the median (Med), mode (Mod) and standard deviation (Std) for the posterior distribution of each parameter is shown, as well as the 68.3\% (CI(15.85),CI(84.15)) and 95.45\% (CI(2.275),CI(97.725)) confidence intervals. The prior for each parameter can be of type: $\mathcal{U}$: uniform, $\mathcal{N}$: normal, $\mathcal{SN}$:split normal, $\mathcal{TN}$: truncated normal.}
\end{center}
\end{table*}

\begin{table*}
\caption{Parameters probed by the MCMC used to fit the RV measurements of Gl617A - 2-keplerian model plus quadratic drift.}   \label{Gl617A_tab-mcmc-Probed_params_k2d2}
\tiny
\begin{center}
\begin{tabular}{lcclcccccccc}
\hline
\hline
Param. & Units & Max(Like) & Med & Mod &Std & CI(15.85) & CI(84.15) &CI(2.275) & CI(97.725) & Prior\\
\hline
\multicolumn{11}{c}{ \bf Likelihood}\\
\hline
$\log{(\rm Post})$&               &  -434.164980&  -440.075963&  -440.047078&     2.381702&  -443.144186&  -437.799233&  -446.780069&  -436.227907&               \\ 
$\log{(\rm Like)}$&               &  -433.765721&  -439.525314&  -439.289290&     2.390269&  -442.596771&  -437.243499&  -446.189031&  -435.601729&               \\ 
$\log{(\rm Prior)}$&               &    -0.399259&    -0.486225&    -0.368662&     0.298923&    -0.882288&    -0.243982&    -1.520647&    -0.083642&               \\ 
\hline 
M$_{\star}$    &[M$_{\odot}$]  &     0.581188&     0.598699&     0.589123&     0.026717&     0.569082&     0.629830&     0.538640&     0.660030&$\mathcal{U}$  \\ 
\hline 
$\sigma_{JIT}$ &[m\,s$^{-1}$]  &     3.47&     3.71&     3.83&     0.42&     3.25&     3.22&     2.86&     2.68&$\mathcal{U}$  \\ 
\hline 
$\gamma_{SOPHIE}$&[m\,s$^{-1}$]  &-18714.90&-18715.26&-18716.78&     4.92&-18720.89&-18709.68&-18726.32&-18703.97&$\mathcal{U}$  \\ 
\hline 
$lin$          &[m\,s$^{-1}$\,yr$^{-1}$]&    -3.13&    -2.98&    -2.46&     2.19&    -5.47&    -0.48&    -8.10&     1.92&$\mathcal{U}$  \\ 
$quad$         &[m\,s$^{-1}$\,yr$^{-2}$]&     0.32&     0.30&     0.24&     0.23&     0.04&     0.57&    -0.22&     0.85&$\mathcal{U}$  \\ 
\hline 
$P$            &[d]            &    86.685971&    86.716536&    86.644378&     0.163173&    86.533594&    86.911495&    86.364597&    87.082150&$\mathcal{U}$  \\ 
$K$            &[m\,s$^{-1}$]  &     6.51&     6.57&     6.62&     0.33&     6.19&     6.93&     5.83&     7.32&$\mathcal{U}$  \\ 
$e$            &               &     0.232050&     0.230947&     0.204544&     0.066766&     0.152388&     0.303804&     0.070509&     0.372129&$\mathcal{U}$  \\ 
$\omega$       &[deg]          &    91.237907&    97.248455&    96.102520&    12.239918&    83.790120&   110.796187&    66.715021&   127.067212&$\mathcal{U}$  \\ 
$\lambda_{0}$  &[deg]          &   215.291405&   220.718337&   216.160185&    16.196583&   202.266759&   239.462980&   185.195600&   257.350058&$\mathcal{U}$  \\ 
\hline 
$P$            &[d]            &   485.429776&   496.902333&   484.032769&    26.820241&   475.087228&   532.349347&   456.095460&   579.472904&$\mathcal{U}$  \\ 
$K$            &[m\,s$^{-1}$]  &     3.38&     3.16&     3.03&     0.37&     2.74&     3.59&     2.33&     4.03&$\mathcal{U}$  \\ 
$e$            &               &     0.134236&     0.146812&     0.021121&     0.110116&     0.043096&     0.298966&     0.005614&     0.458703&$\mathcal{U}$  \\ 
$\omega$       &[deg]          &  -358.202334&  -311.969148&  -380.751216&   423.482663&  -375.233350&   668.453999&  -478.847476&   714.163530&$\mathcal{U}$  \\ 
$\lambda_{0}$  &[deg]          &   240.692904&   274.875882&   256.683399&    77.426560&   207.978823&   379.031065&   146.361972&   491.545801&$\mathcal{U}$  \\ 
\hline 
\end{tabular}
\tablefoot{The maximum likelihood solution (Max(Like)), the median (Med), mode (Mod) and standard deviation (Std) of the posterior distribution for each parameter is shown, as well as the 68.3\% (CI(15.85),CI(84.15)) and 95.45\% (CI(2.275),CI(97.725)) confidence intervals. The prior for each parameter can be of type: $\mathcal{U}$: uniform, $\mathcal{N}$: normal, $\mathcal{SN}$:split normal, $\mathcal{TN}$: truncated normal.}
\end{center}
\end{table*}

\begin{table*}
\caption{Physical parameters derived from the MCMC chains used fit the RV measurements of Gl617-A - 2-keplerian model plus quadratic drift.}  \label{Gl617A_tab-mcmc-Physical_params_k2d2}
\tiny
\begin{center}
\begin{tabular}{lcclcccccccc}
\hline
\hline
Param. & Units & Max(Like) & Med & Mod &Std & CI(15.85) & CI(84.15) &CI(2.275) & CI(97.725) & Prior\\
\hline
\multicolumn{10}{c}{ \bf Likelihood}\\
\hline
M$_{\star}$    &[M$_{\odot}$]  &     0.581188&     0.598699&     0.589123&     0.026717&     0.569082&     0.629830&     0.538640&     0.660030&$\mathcal{U}$  \\ 
\hline 
$P$            &[d]            &    86.685971&    86.716536&    86.644378&     0.163173&    86.533594&    86.911495&    86.364597&    87.082150&$\mathcal{U}$  \\ 
$K$            &[m\,s$^{-1}$]  &     6.51&     6.57&     6.62&     0.33&     6.19&     6.93&     5.83&     7.32&$\mathcal{U}$  \\ 
$e$            &               &     0.232050&     0.230947&     0.204544&     0.066766&     0.152388&     0.303804&     0.070509&     0.372129&$\mathcal{U}$  \\ 
$\omega$       &[deg]          &    91.237907&    97.248455&    96.102520&    12.239918&    83.790120&   110.796187&    66.715021&   127.067212&$\mathcal{U}$  \\ 
$T_P$          &[d]            & 55470.128617& 55470.170612& 55469.066335&     4.360219& 55465.320275& 55475.139385& 55459.955671& 55480.110606&               \\ 
$T_C$          &[d]            & 55469.947886& 55469.184485& 55469.744721&     3.803915& 55464.775258& 55473.489374& 55460.507190& 55477.410804&               \\ 
\hline 
$Ar$           &[AU]           &     0.319911&     0.323187&     0.321331&     0.004833&     0.317754&     0.328762&     0.311972&     0.333917&               \\ 
M.$\sin{i}$    &[M$_{\rm Jup}$]&     0.096018&     0.098450&     0.097920&     0.005996&     0.091680&     0.105364&     0.085691&     0.113023&               \\ 
M.$\sin{i}$    &[M$_{\rm Earth}$]&    30.514385&    31.287511&    31.118932&     1.905555&    29.135990&    33.484633&    27.232456&    35.918729&               \\ 
\hline 
$P$            &[d]            &   485.429776&   496.902333&   484.032769&    26.820241&   475.087228&   532.349347&   456.095460&   579.472904&$\mathcal{U}$  \\ 
$K$            &[m\,s$^{-1}$]  &     3.38&     3.16&     3.03&     0.37&     2.74&     3.59&     2.33&     4.03&$\mathcal{U}$  \\ 
$e$            &               &     0.134236&     0.146812&     0.021121&     0.110116&     0.043096&     0.298966&     0.005614&     0.458703&$\mathcal{U}$  \\ 
$\omega$       &[deg]          &  -358.202334&  -311.969148&  -380.751216&   423.482663&  -375.233350&   668.453999&  -478.847476&   714.163530&$\mathcal{U}$  \\ 
$T_P$          &[d]            & 55177.869828& 55214.752874& 55283.022634&   135.078647& 55031.879715& 55361.011891& 54952.197623& 55438.122406&               \\ 
$T_C$          &[d]            & 55276.199082& 55275.541145& 55277.491526&   107.612883& 55154.204758& 55385.387258& 55055.822719& 55543.038427&               \\ 
\hline 
$Ar$           &[AU]           &     1.008817&     1.035866&     1.026245&     0.039843&     0.999836&     1.087041&     0.965852&     1.152976&               \\ 
M.$\sin{i}$    &[M$_{\rm Jup}$]&     0.090084&     0.085769&     0.082725&     0.010332&     0.074076&     0.097851&     0.062734&     0.109579&               \\ 
M.$\sin{i}$    &[M$_{\rm Earth}$]&    28.628775&    27.257285&    26.290082&     3.283456&    23.541287&    31.097041&    19.936952&    34.824225&               \\ \hline 
\end{tabular}
\tablefoot{The maximum likelihood solution (Max(Like)), the median (Med), mode (Mod) and standard deviation (Std) for the posterior distribution of each parameter is shown, as well as the 68.3\% (CI(15.85),CI(84.15)) and 95.45\% (CI(2.275),CI(97.725)) confidence intervals. The prior for each parameter can be of type: $\mathcal{U}$: uniform, $\mathcal{N}$: normal, $\mathcal{SN}$:split normal, $\mathcal{TN}$: truncated normal.}
\end{center}
\end{table*}

\begin{table*}
\caption{Parameters probed by the MCMC used to fit the combined SOPHIE, CARMENES, and KECK RV measurements of Gl617A.} \label{Gl617A_alldat_tab-mcmc-Probed_params}
\tiny
\begin{center}
\begin{tabular}{lcclcccccccc}
\hline
\hline
Param. & Units & Max(Like) & Med & Mod &Std & CI(15.85) & CI(84.15) &CI(2.275) & CI(97.725) & Prior\\
\hline
\multicolumn{11}{c}{ \bf Likelihood}\\
\hline
$\log{(\rm Post})$&               &  -934.737509&  -939.951019&  -940.359405&     2.181801&  -942.707691&  -937.845539&  -946.160275&  -936.299385&               \\ 
$\log{(\rm Like)}$&               &  -931.716450&  -937.193692&  -936.740517&     2.633389&  -940.751319&  -934.848899&  -944.875654&  -933.276083&               \\ 
$\log{(\rm Prior)}$&               &    -3.021059&    -3.015167&    -3.129632&     1.055714&    -3.056399&    -0.160778&    -3.122142&    -0.006795&               \\ 
\hline 
M$_{\star}$    &[M$_{\odot}$]  &     0.617017&     0.599486&     0.595975&     0.025964&     0.569661&     0.628938&     0.539536&     0.658046&$\mathcal{U}$  \\ 
\hline 
$\sigma_{JIT}$ &[m\,s$^{-1}$]  &    14.15&    15.89&    15.31&     2.25&    14.07&    20.03&    12.57&    21.49&$\mathcal{U}$  \\ 
\hline 
$\gamma_{CARMENES}$&[m\,s$^{-1}$]  &     0.51&     0.40&     0.14&     1.44&    -1.20&     2.03&    -2.85&     3.80&$\mathcal{U}$  \\ 
$\gamma_{KECK-PUB_1}$&[m\,s$^{-1}$]  &    -2.55&    -1.76&    -1.60&     2.37&    -4.44&     0.87&    -7.25&     3.72&$\mathcal{U}$  \\ 
$\gamma_{SOPHIE}$&[m\,s$^{-1}$]  &-18721.42&-18721.59&-18721.85&     1.34&-18723.09&-18720.05&-18724.63&-18718.48&$\mathcal{U}$  \\ 
\hline 
$lin$          &[m\,s$^{-1}$\,yr$^{-1}$]&    -0.20&    -0.12&    -0.21&     0.40&    -0.57&     0.32&    -1.06&     0.79&$\mathcal{U}$  \\ 
$quad$         &[m\,s$^{-1}$\,yr$^{-2}$]&     0.01&    -0.00&    -0.02&     0.04&    -0.05&     0.04&    -0.09&     0.09&$\mathcal{U}$  \\ 
\hline 
$P$            &[d]            &    86.717889&    86.776472&    86.776265&     0.135383&    86.630677&    86.933133&    86.478695&    87.110679&$\mathcal{U}$  \\ 
$K$            &[m\,s$^{-1}$]  &     5.77&     5.83&     5.76&     0.20&     5.59&     6.05&     5.38&     6.28&$\mathcal{U}$  \\ 
$e$            &               &     0.067519&     0.071525&     0.072428&     0.035663&     0.030243&     0.112694&     0.004864&     0.159102&$\mathcal{U}$  \\ 
$\omega$       &[deg]          &    74.887523&    97.219922&    98.412412&    34.567998&    55.385453&   127.614449&   -16.938101&   176.384852&$\mathcal{U}$  \\ 
$\lambda_{0}$  &[deg]          &   220.983105&   227.841194&   223.343831&    14.690130&   211.859018&   244.846508&   195.405489&   263.911937&$\mathcal{U}$  \\ 
\hline 
\end{tabular}
\tablefoot{The maximum likelihood solution (Max(Like)), the median (Med), mode (Mod) and standard deviation (Std) for the posterior distribution of each parameter is shown, as well as the 68.3\% (CI(15.85),CI(84.15)) and 95.45\% (CI(2.275),CI(97.725)) confidence intervals. The prior for each parameter can be of type: $\mathcal{U}$: uniform, $\mathcal{N}$: normal, $\mathcal{SN}$:split normal, $\mathcal{TN}$: truncated normal.}
\end{center}
\end{table*}

\begin{table*}
\scriptsize
\caption{Physical parameters derived from the MCMC chains used fit the combined SOPHIE, CARMENES, and KECK RV measurements of Gl617A.}.  \label{Gl617A_alldat_tab-mcmc-Physical_params}
\def\arraystretch{1.2}
\begin{center}
\begin{tabular}{lcclcccccccc}
\hline
\hline
Param. & Units & Max(Like) & Med & Mod &Std & CI(15.85) & CI(84.15) &CI(2.275) & CI(97.725) & Prior\\
\hline
\multicolumn{10}{c}{ \bf Likelihood}\\
\hline
M$_{\star}$    &[M$_{\odot}$]  &     0.617017&     0.599486&     0.595975&     0.025964&     0.569661&     0.628938&     0.539536&     0.658046&$\mathcal{U}$  \\ 
\hline 
$P$            &[d]            &    86.717889&    86.776472&    86.776265&     0.135383&    86.630677&    86.933133&    86.478695&    87.110679&$\mathcal{U}$  \\ 
$K$            &[m\,s$^{-1}$]  &     5.77&     5.83&     5.76&     0.20&     5.59&     6.05&     5.38&     6.28&$\mathcal{U}$  \\ 
$e$            &               &     0.067519&     0.071525&     0.072428&     0.035663&     0.030243&     0.112694&     0.004864&     0.159102&$\mathcal{U}$  \\ 
$\omega$       &[deg]          &    74.887523&    97.219922&    98.412412&    34.567998&    55.385453&   127.614449&   -16.938101&   176.384852&$\mathcal{U}$  \\ 
$T_P$          &[d]            & 55464.808054& 55468.264643& 55469.553885&     9.043231& 55457.599658& 55476.783117& 55440.927623& 55489.460498&               \\ 
$T_C$          &[d]            & 55467.985292& 55466.914469& 55465.796807&     3.657568& 55462.799889& 55471.033793& 55458.170838& 55475.281484&               \\ 
\hline 
$Ar$           &[AU]           &     0.326432&     0.323420&     0.322034&     0.004684&     0.317906&     0.328658&     0.312283&     0.333674&               \\ 
M.$\sin{i}$    &[M$_{\rm Jup}$]&     0.090837&     0.089847&     0.090072&     0.004054&     0.085275&     0.094534&     0.080783&     0.099046&               \\ 
M.$\sin{i}$    &[M$_{\rm Earth}$]&    28.867969&    28.553526&    28.624925&     1.288239&    27.100269&    30.042914&    25.672842&    31.476762&               \\ 
\hline 
\end{tabular}
\tablefoot{The maximum likelihood solution (Max(Like)), the median (Med), mode (Mod) and standard deviation (Std) for the posterior distribution of each parameter is shown, as well as the 68.3\% (CI(15.85),CI(84.15)) and 95.45\% (CI(2.275),CI(97.725)) confidence intervals. The prior for each parameter can be of type: $\mathcal{U}$: uniform, $\mathcal{N}$: normal, $\mathcal{SN}$:split normal, $\mathcal{TN}$: truncated normal.}
\end{center}
\end{table*}

\end{appendix}

\end{document}